\def\Ar{\rightarrow}
\def\bar{\overline}

\def\b{\beta}
\def\n{\nu}
\def\m{\mu}

\def\ra{\rightarrow}
\def\t{\tilde}
\def\bar{\overline}
\def\l{\lambda}
\def\G{{\rm GeV}}

\def\eV{{\rm eV}}
\def\sA{{\scriptscriptstyle A }}
\def\sB{{\scriptscriptstyle B }}
\def\sC{{\scriptscriptstyle C }}

\def\sL{{\scriptscriptstyle L }}
\def\sM{{\scriptscriptstyle M }}
\def\sN{{\scriptscriptstyle N }}

\def\sQ{{\scriptscriptstyle Q }}
\def\sR{{\scriptscriptstyle R }}

\def\sS{{\scriptscriptstyle S }}

\def\sW{{\scriptscriptstyle W }}
\def\sX{{\scriptscriptstyle X }}
\def\sZ{{\scriptscriptstyle Z }}

\def\diag{{\scriptscriptstyle\rm diag}}

\def\>{{\stackrel{>}{\scriptstyle\sim}}}
\def\<{{\stackrel{<}{\scriptstyle\sim}}}

\def\MNS{{\rm \sM\sN\sS}}

\documentstyle [12pt,epsf]{article}
\clearpage
\makeatletter

\@addtoreset{equation}{section}
\makeatother
\begin{document}
\baselineskip=24pt
\setcounter{page}{1}
\thispagestyle{empty}
\topskip 2.5  cm
\topskip 0.5  cm
\centerline{\LARGE \bf Lepton Flavor Violating Processes} 
\vskip 0.5 cm
\centerline{\LARGE \bf in Bi-maximal Texture of Neutrino Mixings} 
\vskip 0.7 cm
\vskip 1.0 cm
\centerline{{\large \bf Atsushi Kageyama}
 \renewcommand{\thefootnote}{\fnsymbol{footnote}}
\footnote[1]{E-mail address:  atsushi@muse.hep.sc.niigata-u.ac.jp},
\qquad{\large \bf Satoru Kaneko}
\renewcommand{\thefootnote}{\fnsymbol{footnote}}
\footnote[2]{E-mail address: kaneko@muse.hep.sc.niigata-u.ac.jp}}
\vskip 0.5 cm
\centerline{{\large \bf Noriyuki Shimoyama}
\renewcommand{\thefootnote}{\fnsymbol{footnote}}
\footnote[3]{E-mail address: simoyama@muse.hep.sc.niigata-u.ac.jp},
\qquad {\large \bf Morimitsu Tanimoto}
\renewcommand{\thefootnote}{\fnsymbol{footnote}}
\footnote[4]{E-mail address: tanimoto@muse.hep.sc.niigata-u.ac.jp}
 }
\vskip 0.5 cm
 \centerline{ \it{Department of Physics, Niigata University, 
 Ikarashi 2-8050, 950-2181 Niigata, JAPAN}}
\vskip 1.0  cm
\centerline{\bf ABSTRACT}\par
\vskip 0.3 cm

We investigate the lepton flavor violation in the framework of the MSSM with
right-handed neutrinos taking the large mixing angle MSW solution in the 
quasi-degenerate and  the inverse-hierarchical neutrino masses.
We predict the branching ratio of $\mu \rightarrow
e+\gamma$ and $\tau \rightarrow \mu+\gamma$ processes assuming the degenerate
right-handed Majorana neutrino masses.  We find that the branching ratio
in the quasi-degenerate neutrino mass spectrum is 100 times smaller
than the ones in the inverse-hierarchical 
and the hierarchical neutrino spectra.
We emphasize  that  the magnitude of  $U_{e3}$ is one of  important
ingredients to predict BR($\mu \rightarrow e +\gamma $).
The effect  of the deviation from the complete-degenerate
 right-handed Majorana neutrino masses are also estimated.
Furtheremore, we examine  the $S_{3\sL}\times S_{3\sR}$ model, which gives
the quasi-degenerate neutrino masses,  and the Shafi-Tavartkiladze model,
which gives the inverse-hierarchical neutrino masses.
Both predicted branching ratios of $\mu\rightarrow e+\gamma$ 
are smaller than the experimantal bound.
 
\newpage
\topskip 0. cm
\section{Introduction}
 Super-Kamiokande  has almost confirmed  the neutrino oscillation
 in the atmospheric neutrinos, which favors the $\n_\mu\Ar \nu_\tau$
process  \cite{SKam}.
For the solar neutrinos \cite{SKamsolar,SNO}, the recent data of 
the Super-Kamiokande and the SNO also suggest the neutrino oscillation
$\n_e\Ar \nu_{\rm x}$ with the large mixing angle (LMA) MSW solution,  
although other solutions are still allowed \cite{MSW,Lisi}.
If we take the LMA-MSW solution,  neutrinos are  massive and 
the flavor mixings are almost  bi-maximal  in the lepton sector. 
 
If neutrinos are massive and mixed in the SM, 
there exists a source of the lepton flavor violation (LFV) through the
off-diagonal elements of the neutrino Yukawa coupling matrix. 
However, due to the smallness of the neutrino masses, 
the predicted branching ratios for these processes are so tiny that they 
are completely unobservable\cite{SM}.


On the other hand,
in the supersymmetric framework  the situation is quite different.
Many authors have already studied the LFV  in the minimal supersymmetric standard model (MSSM) with right-handed neutrinos assuming the relevant neutrino 
mass matrix \cite{LFV1,LFV2,Sato,Casas}.
In the MSSM with soft breaking terms, 
there exist  lepton flavor violating terms such as 
off-diagonal elements of slepton mass matrices 
$\left({\bf m}^2_{\tilde \sL}\right)_{ij}$,
$({\bf m}_{ \t e_\sR}^2)_{ij}$ 
and trilinear couplings ${\bf A}^{e}_{ij}$.
Strong bounds on these matrix elements come from requiring branching ratios 
for LFV processes to be below observed ratios.  
For the present, the most stringent bound comes from  the 
$\mu \rightarrow e +\gamma $ decay 
(${\rm BR}(\m\ra e+\gamma)<1.2\times 10^{-11}$) \cite{MuEGamma}.
However, if the LFV  occurs at tree level in the soft breaking terms, 
the branching ratio of this process exceeds the experimental bound
considerably.
Therefore  one assumes that the LFV does not occur at tree level 
in the soft parameters. 
This is realized by taking the assumption that soft parameters 
such as $\left({\bf m}^2_{\t \sL}\right)_{ij}$,
$({\bf m}_{\t e_\sR}^2 )_{ij}$,  ${\bf A}^{e}_{ij}$,
are universal {\it i.e.}, proportional to the unit matrix. 
This assumption follows from the minimal supergravity (m-SUGRA).  
However, even though there is no flavor violation at tree level, 
it is generated by the effect of
the renormalization group equations (RGE's) via neutrino Yukawa couplings. 
Suppose that  neutrino masses are produced by the  see-saw mechanism 
\cite{seesaw}, there are  the right-handed neutrinos 
above a  scale $M_{\sR}$.
Then neutrinos 
have  the Yukawa coupling matrix ${\bf Y}_\nu$  with  off-diagonal entries
in the basis of the diagonal charged-lepton Yukawa couplings.
 The  off-diagonal elements of ${\bf Y}_\nu$ drive off-diagonal ones 
in the $\left({\bf m}_{\rm\t \sL}^2\right)_{ij}$ and 
${\bf A}^{e}_{ij}$ matrices through the RGE's running \cite{Borzumati}.

One can construct  ${\bf Y}_\nu$ by the recent data of neutrino oscillations.
   Assuming that oscillations need only accounting for 
 the solar and the atmospheric neutrino data, we take  the 
LMA-MSW solution of the solar neutrino.  
Then, the lepton mixing matrix,
 which may be called  the MNS matrix or the MNSP  matrix \cite{MNS,Po}, 
is given in ref.\cite{FT}. 
Since the data of neutrino oscillations only indicate
the differences of the mass square   $\Delta m^2_{ij}$,
  neutrinos have three possible mass spectra:  the hierarchical spectrum
  $m_{\n 3}\gg  m_{\n 2} \gg  m_{\n 1}$ , the quasi-degenerate one
  $m_{\n 1}\simeq m_{\n 2}\simeq m_{\n 3}$ 
  and the  inverse-hierarchical one  $m_{\n 1}\simeq m_{\n 2} \gg  m_{\n 3}$.

We have already analyzed the effect of neutrino Yukawa couplings
for the  $\mu \rightarrow e +\gamma$  process assuming  
the quasi-degenerate one and the inverse-hierarchical one \cite{our}.
In this paper, we present  detailed formula in our calculations of
$\mu \rightarrow e +\gamma$ and discuss the dependence of the SUSY breaking 
parameters for the branching ratio.
In the previous paper, the right-handed Majorana neutrino masses are
assumed to be completely degenerate. We study  the effect of the
deviation from this degeneracy in this work.
The correlation between 
BR($\mu \rightarrow e +\gamma $) and  BR($\tau \rightarrow \mu +\gamma $)
is also calculated.
Furtheremore, two specific models of the neutrino mass matrix are examined 
in the  $\mu \rightarrow e +\gamma $ process.

This paper is organized as follows. In section 2, we give the general form
of ${\bf Y}_\n$ and ${\bf Y}_\n^{\dagger}{\bf Y}_\n$, which play a
crucial role in generating the LFV through the RGE's running. 
In section 3, we calculate the branching ratio of the processes 
$\mu \ra e+\gamma$ and $\tau \ra \mu+\gamma$, respectively in the three 
neutrino mass spectra. 
In section 4, we examine the $S_{3\sL}\times S_{3\sR}$ model, which gives the
quasi-degenerate neutrino masses,  and the Shafi-Tavartkiladze model,
which gives the inverse-hierarchical neutrino masses.
In section 5, we summarize our results and give
discussions.

\section{LFV in the MSSM with Right-handed Neutrinos}
\subsection{Yukawa Couplings}
In this section, 
we introduce the general expression of the neutrino Yukawa coupling 
${\bf Y}_\n$, which is useful in the following arguments, 
and investigate the LFV triggered by the neutrino Yukawa couplings.
The superpotential of the lepton sector is described as follows:
\begin{eqnarray}
 W_{\rm lepton}
&=&
{\bf Y}_e LH_de_{\sR}^c+{\bf Y_\n} LH_u\n_{\sR}^c
+\frac{1}{2}{\n_\sR^c}^T {\bf M_\sR}\n_\sR^c \ ,
\end{eqnarray}
where $H_u, H_d$ are chiral superfields for Higgs doublets, $L$ is
the left-handed lepton doublet, $e_\sR$ and $\n_\sR$ are the right-handed
charged lepton 
and the neutrino superfields, respectively. The ${\bf Y}_e$ 
is the Yukawa coupling matrix for the charged lepton,
${\bf M_\sR}$ is Majorana mass matrix of the right-handed neutrinos. We take
${\bf Y}_e$ and ${\bf M_\sR}$ to be diagonal.

It is well-known that the neutrino mass matrix is given as 
\begin{eqnarray}
 {\bf m_{\n}}=
\left({\bf Y_\n} v_u\right)^{\rm T}{\bf M}_{\bf \sR}^{-1}
\left({\bf Y_\n} v_u\right)\ ,
\label{eqn:mn}
\end{eqnarray}
via the see-saw mechanism, where $v_u$ is the vacuum
expectation value (VEV) of Higgs $H_u$. 
In eq.(\ref{eqn:mn}), the Majorana mass term for left-handed neutrinos is not
included since we consider the minimal extension of the MSSM.

 The neutrino mass matrix ${\bf m_\n}$ is diagonalized by a single unitary
 matrix
\begin{eqnarray}
{\bf m_\n^{\diag}}
&\equiv&
{\bf U}_{\MNS}^{\rm T}{\bf m_\n} {\bf U_{\MNS}} \ ,
\label{eqn:mndiag}
\end{eqnarray}
where ${\bf U_{\MNS}}$ is the lepton mixing matrix. 
Following the expression in ref.\cite{Casas}, we write the neutrino Yukawa
 coupling as 
\begin{eqnarray}
{\bf Y_\n}
=
\frac{1}{v_u}
\sqrt{{\bf M_\sR^{\diag}}}\ {\bf R}\ \sqrt{{\bf m_\n^{\diag}}} \ 
{\bf U}^{\rm T}_{\MNS} \ ,
\end{eqnarray}
or explicitly
\begin{equation}
{\bf Y_\nu}
= \frac{1}{v_u}  \left (\matrix{\sqrt{M_{\sR 1}}& 0 & 0\cr
  0 & \sqrt{M_{\sR 2}}  & 0 \cr  0 & 0 & \sqrt{M_{\sR 3}} \cr  } \right) 
  {\bf R}
  \left (\matrix{\sqrt{m_{\n 1}}& 0 & 0\cr
  0 & \sqrt{m_{\n 2}}  & 0 \cr  0 & 0 & \sqrt{m_{\n 3}} \cr  } \right)  
 {\bf U}^{\rm T}_{\MNS}\ ,
\label{YEW}
\end{equation}
\noindent 
where $\bf R$ is a $3\times 3$ orthogonal matrix, which depends on models.
Details are given in Appendix A.


At first, let us take the degenerate right-handed Majorana
masses  $M_{\sR1}=M_{\sR2}=M_{\sR3}\equiv M_R$.  
This assumption is reasonable for the case of the quasi-degenerate
neutrino masses.
Otherwise a big conspiracy would be needed between ${\bf Y_\nu}$ 
and ${\bf M}_{\bf\sR}$.  
This assumption is also taken for cases of the inverse-hierarchical and
the hierarchical neutrino masses.
We also discuss later the effect of the deviation from the degenerate 
right-handed Majonara neutrino masses.

Then,  we get
\begin{eqnarray}
 {\bf Y_\nu}
= \frac{\sqrt{M_{\sR}}}{v_u} \  {\bf R} \ 
  \left (\matrix{\sqrt{m_{\n 1}}& 0 & 0\cr
  0 & \sqrt{m_{\n 2}}  & 0 \cr  0 & 0 & \sqrt{m_{\n 3}} \cr  } \right)  
{\bf U}^{\rm T}_{\MNS}\ , 
\label{yukawa}
\end{eqnarray}
\noindent and
\begin{eqnarray}
 {\bf Y_\n^{\dagger}Y_\n}
&=&
\frac{M_\sR}{v_u^2}
{\bf U_{\MNS}}
\left (
\matrix{
 m_{\n 1}& 0 & 0\cr
 0 & m_{\n 2}& 0\cr 
 0 & 0 &m_{\n 3}\cr
}
\right)  
{\bf U}_{\MNS}^{\rm T}\ ,
\label{yy}
\end{eqnarray}
or equivalently,
\begin{eqnarray}
 \left({\bf Y_\n^{\dagger}Y_\n}\right)_{\alpha\beta}
&=&
\frac{M_\sR}{v_u^2}\sum_{i=1}^{3}m_{\n i}U_{\alpha i}U_{\beta i}^*\ ,
\end{eqnarray} 
where $U_{\alpha\beta}$'s are the elements of ${\bf U_{\MNS}}$.
It is remarked that $\bf Y_\n^{\dagger}Y_\n$ is independent of
 $\bf R$ in the case of  $M_{\sR1}=M_{\sR2}=M_{\sR3}\equiv M_\sR$.
It may be important to consider the deviation from the degenerate 
right-handed Majonara neutrino masses. Detailed discussions are given
 in subsection 3.2.

Note that this representation of the Yukawa coupling is given at 
the electroweak scale.
Since we need the Yukawa coupling at the GUT scale, 
 eq.(\ref{YEW}) should be modified by taking account of the effect of 
the RGE's \cite{RGE1,RGE2,Haba}.
Modified Yukawa couplings at a scale $M_\sR$   are given as 
\begin{equation}
{\bf Y_\nu}
= \frac{\sqrt{M_\sR}}{v_{u}}\  {\bf R}\  \left (\matrix{\sqrt{m_{\n 1}}& 0 & 0\cr
  0 & \sqrt{m_{\n 2}}  & 0 \cr  0 & 0 & \sqrt{m_{\n 3}} \cr  } \right)  
{\bf U}^{\rm T}_{\MNS} \sqrt{I_g I_t}
\left (\matrix{1& 0 & 0\cr 0 & 1 & 0 \cr 0 & 0 & \sqrt{I_\tau}\cr}\right)
\ ,
\label{YGUT}
\end{equation}
with
\begin{equation}
I_g =\exp\left [\frac{1}{8\pi^2}\int_{t_\sZ}^{t_\sR} -c_i g_i^2 dt\right ]
\ ,\quad
I_t =\exp \left [\frac{1}{8\pi^2}\int_{t_\sZ}^{t_\sR} y_t^2 dt \right ] 
\ ,  \quad
I_{\tau}= \exp\left [\frac 1{8\pi^2}\int_{t_\sZ}^{t_\sR} y_{\tau}^2 dt 
\right ]\ ,
\end{equation}
where  $t_\sR=\ln M_\sR$ and $t_\sZ=\ln M_\sZ$. Here, $g_i$'s$(i=1,2)$ are 
gauge couplings and  $y_t$ and $y_\tau$ are Yukawa couplings,  
$c_i$'s are the constants $( \frac{3}{5},3 )$. 
We shall calculate the LFV numerically by using the modified Yukawa
coupling in the following sections.

As mentioned in the previous section, there are three possible neutrino
mass spectra.
The hierarchical type ($m_{\n 1}\ll m_{\n 2}\ll m_{\n 3}$) 
gives the neutrino mass spectrum as
\begin{eqnarray}
 m_{\n 1}\sim 0 \  ,\qquad m_{\n 2}=\sqrt{\Delta m^2_{\odot}} \  ,
  \qquad m_{\n 3}=\sqrt{\Delta m^2_{\rm atm}}\ ,
\end{eqnarray}
the quasi-degenerate type ($m_{\n 1}\sim m_{\n 2}\sim m_{\n 3}$) gives 
\begin{equation}
 m_{\n 1}\equiv m_\nu \ ,  
 \qquad  m_{\n 2}=m_\nu + \frac{1}{2 m_\nu}\Delta m^2_{\odot}\  , \qquad
 m_{\n 3}=m_\nu +  \frac{1}{2 m_\nu} \Delta m_{\rm atm}^2 \ , 
\label{eqn:degeneratetype}
\end{equation}
and the inverse-hierarchical type ($m_{\n 1}\sim m_{\n 2}\gg m_{\n 3}$) gives
\begin{equation}
 \quad  m_{\n 2}
\equiv
\sqrt{\Delta m^2_{\rm atm}} \  , \quad
 m_{\n 1}
=
 m_{\n 2}-\frac{1}{2 m_{\n 2}}\Delta m_{\odot}^2 \ , \quad  
 m_{\n 3}\simeq 0\ .
\end{equation}
We take the typical values 
$\Delta m^2_{\rm atm}=  3\times  10^{-3} \eV^2$ and 
$\Delta m_{\odot}^2= 7 \times  10^{-5}  \eV^2$ 
in our calculation of the LFV.

We take the typical mixing angles of the LMA-MSW solution such as 
$s_{23}=1/\sqrt{2}$ and $s_{12}=0.6$ \cite{FT}, in which  
  the lepton mixing matrix is given in  terms of the standard
  parametrization of 
the mixing matrix \cite{PDG} as follows:
\begin{equation}  
{\bf U_{\MNS}}
=  \left (\matrix{ c_{13} c_{12} & c_{13} s_{12} &  s_{13} e^{-i \phi}\cr 
  -c_{23}s_{12}-s_{23}s_{13}c_{12}e^{i \phi} & 
c_{23}c_{12}-s_{23}s_{13}s_{12}e^{i \phi} &   s_{23}c_{13} \cr
  s_{23}s_{12}-c_{23}s_{13}c_{12}e^{i \phi} & 
 -s_{23}c_{12}-c_{23}s_{13}s_{12}e^{i \phi} &  c_{23}c_{13} \cr} \right ) \ ,
\end{equation} 
where  $s_{ij}\equiv \sin{\theta_{ij}}$ and 
 $c_{ij}\equiv \cos{\theta_{ij}}$ are mixings in vacuum, 
 and $\phi$ is the $CP$ violating phase.  
The reacter experiment of  CHOOZ  \cite{Chooz} presented a upper bound of 
$s_{13}$.  We use the  constraint from the two flavor analysis, which is 
$s_{13}\leq 0.2$ in our calculation.
If we take account of  the recent result of the three flavor analysis
\cite{Three},  the upper bound of $s_{13}$ may be smaller than 0.2.
Then,  if we use the results in \cite{Three}, our results  of $\mu \rightarrow e+\gamma$ are reduced at most by a factor of two. 
In our calculation, the  CP violating phase is neglected for simplicity.

\subsection{LFV in Slepton Masses}

Since SUSY is spontaneously broken at the low energy, 
we consider the MSSM with the soft SUSY breaking terms:
\begin{eqnarray}
-{\cal{L}}_{\rm soft}
&=&
({\bf m}_{\tilde \sQ}^2)_{ij} {\tilde Q}_{i}^{\dagger}{\tilde Q}_{j}
+({\bf m}_{\tilde u}^2)_{ij} {\tilde u}_{\sR i}^* {\tilde u}_{\sR j}
+({\bf m}_{\tilde d}^2)_{ij} {\tilde d}_{\sR i}^* {\tilde d}_{\sR j}
\nonumber \\
& &
+({\bf m}_{\tilde \sL}^2)_{ij} {\tilde L}_{i}^{\dagger}{\tilde L}_{j}
+({\bf m}_{\tilde e}^2)_{ij} {\tilde e}_{\sR i}^* {\tilde e}_{\sR j}
+({\bf m}_{\tilde \nu}^2)_{ij} {\tilde \nu}_{\sR i}^* {\tilde \nu}_{\sR j}
\nonumber \\
& &+{\widetilde m}^2_{H_d}H_d^{\dagger} H_d
+{\widetilde m}^2_{H_u}H_u^{\dagger} H_u
+(B \mu H_d H_u
+\frac{1}{2}B_{\nu ij} M_{\sR ij}{\tilde \nu}_{\sR i}^* {\tilde \nu}_{\sR j}^* + h.c.)
\nonumber \\
& &
+ ( {\bf A}^{d}_{ij}  H_d {\tilde d}_{\sR i}^*{\tilde Q}_{j}
   +{\bf A}^{u}_{ij}  H_u {\tilde u}_{\sR i}^*{\tilde Q}_{j}
   +{\bf A}^{e}_{ij}  H_d {\tilde e}_{\sR i}^*{\tilde L}_{j}
   +{\bf A}^{\nu}_ {ij}H_u {\tilde \nu}_{\sR i}^*{\tilde L}_{j}
\nonumber \\
& & 
+\frac{1}{2}M_1 {\tilde B}_L^0 {\tilde B}_L^0
+\frac{1}{2}M_2 {\tilde W}_L^a {\tilde W}_L^a
+\frac{1}{2}M_3 {\tilde G}^a {\tilde G}^a +h.c.)\ ,
\end{eqnarray}
where ${\bf m}_{\tilde Q}^2, {\bf m}_{\tilde u}^2, {\bf m}_{\tilde d}^2, 
{\bf m}_{\tilde \sL}^2, {\bf m}_{\tilde e}^2$ and ${\bf m}_{\tilde \nu}^2$
are mass-squares of the left-handed squark, the right-handed up squark, 
the right-handed down squark, the left-handed charged slepton, 
the right-handed charged slepton and the sneutrino, respectively. 
The ${\widetilde m}^2_{H_d}$ and ${\widetilde m}^2_{H_u}$ are  
mass-squares of Higgs,   
${\bf A}_{d}, {\bf A}_{u}, {\bf A}_{e}$ and ${\bf A}_{\nu}$ are A-parameters
 for squarks and sleptons, and 
$M_1, M_2$ and $M_3$ are the gaugino masses, 
respectively.

Note that the lepton flavor violating processes  come from 
diagrams including  non-zero off-diagonal elements of the soft
parameter.
In this paper we assume the m-SUGRA, therefore we put the assumption 
of universality for soft SUSY breaking terms at the unification scale:
\begin{eqnarray}
({\bf m}_{\tilde \sL}^2)_{ij}
&=&
({\bf m}_{\tilde e}^2)_{ij}
=({\bf m}_{\tilde \nu}^2)_{ij}
=\cdots
=\delta_{ij}m_{0}^2
\ ,
\nonumber\\
{\widetilde m_{H_d}}^2
&=&
{\widetilde m_{H_u}}^2
=m_0^2
\ ,
\nonumber\\
{\bf A}^{\nu}
&=&
{\bf Y}_{\nu}a_0m_0,~
{\bf A}^{e}
={\bf Y}_{e}a_0m_0
\ ,
\nonumber\\
{\bf A}^{u}
&=&
{\bf Y}_{u}a_0m_0,~
{\bf A}^{d}
={\bf Y}_{d}a_0m_0
\ ,
\end{eqnarray}
where  $m_0$ and $a_0$ stand for the universal scalar mass and the universal 
A-parameter, respectively.
Because of the universality, the LFV is not caused
at the unification scale.

To estimate the soft parameters at the low energy,
we need to know the effect of radiative corrections.
As a result, the lepton flavor conservation is violated
at the low energy.

The RGE's  for the left-handed slepton soft mass are given by  
\begin{eqnarray}
&&\mu \frac{d}{d \mu}({\bf m}_{\tilde \sL}^2)_{ij}
=
\left.\mu \frac{d}{d \mu}({\bf m}_{\tilde \sL}^2)_{ij}\right|_{\rm MSSM}
\nonumber\\
&&+
\frac{1}{16 \pi^2}
\left[
({\bf m}_{\tilde \sL}^2 {\bf Y}_{\nu}^{\dagger}{\bf Y}_{\nu}
+{\bf Y}_{\nu}^{\dagger}{\bf Y}_{\nu}{\bf m}_{\tilde \sL}^2)_{ij}
+
2({\bf Y}_{\nu}^{\dagger}{\bf m}_{\tilde \nu}{\bf Y}_{\nu}
+{\tilde m}_{H _{u}}^2 {\bf Y}_{\nu}^{\dagger}{\bf Y}_{\nu}
+{\bf A}_{\nu}^{\dagger}{\bf A}_{\nu})_{ij}
\right] \ ,\nonumber\\
\end{eqnarray}
while  the first term in the right hand side 
is the normal MSSM term which has no LFV, 
and the second one is a source of the LFV through the off-diagonal elements 
of  neutrino Yukawa couplings. 
The RGE's are summarized in Appendix B.

\section{Numerical Analyses of Branching Ratios}

Let us calculate the branching ratio of $e_{i}\rightarrow e_{j}+\gamma\
(j<i)$.
The amplitude of this process is given as
\begin{eqnarray}
 T=e\epsilon^{\alpha*}(q)\bar{u}_{j}(p)m_{e_{i}}i\sigma_{\alpha\beta}q^{\beta}
(A^{\sL}P_{\sL}+A^{\sR}P_{\sR})u_{i}(q-p)\ ,
\label{amplitude}
\end{eqnarray}
where $u_{i}$ is the wave function of $i$-th charged lepton $e_{i}$,  
$p$ and $q$ are momenta of $e_{j}$ and photon, respectively, 
$e$ is the electric charge, 
$\epsilon$ is the polarization vector of photon, and  
$P_{\sL,\sR}$ are  projection operators : $P_{\sL,\sR}=(1\mp\gamma_{5})/2$. 
The $A^{\sL,\sR}$ are decay amplitudes and explicit forms are given in Appendix C.
It is easy to see that this process changes chirality of the charged lepton.
The decay rate can be calculated using $A^{\sL,\sR}$ as 
\begin{eqnarray}
\Gamma(e_{i}\rightarrow e_{j}+\gamma)=
\frac{e^{2}}{16\pi}m_{e_{i}}^{5}
(|A^{\sL}|^{2}+|A^{\sR}|^{2})\ .
\label{rate}
\end{eqnarray}
Since we know the relation $m_{e_{i}}^2\gg m_{e_{j}}^2$, then we can
expect $|A^{\sR}|\gg |A^{\sL}|$.
The $A^{\sL,\sR}$ contain the contribution of the neutralino loop and
the chargino loop as seen in fig.1. 

\begin{figure}
\epsfxsize=10.0 cm
\centerline{\epsfbox{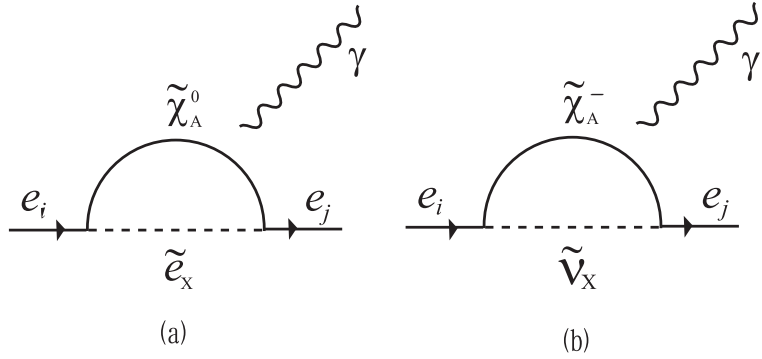}}
\caption{Feynman diagrams which contribute to the branching ratio of
 $e_{i}\rightarrow e_{j}+\gamma$. There are two types of diagrams,
 (a)\,neutralino-slepton loop and (b)\,chargino-sneutrino loop.}
\end{figure}
We calculate the branching ratio using (\ref{rate}) and the formulas in
Appendix C. 
In order to clarify parameter dependence, 
let us present an approximate estimation. 
The decay amplitude is approximated as 
\begin{eqnarray}
 | A^{\sR}|^2 \simeq
\frac{\alpha_{2}^{2}}{16\pi^{2}}
\frac{|(\bigtriangleup {\bf m}_{\tilde{\sL}}^{2})_{ij}|^{2}}{m_{\sS}^{8}}\tan^{2}\beta \ ,
\end{eqnarray}
where  $\alpha_{2}$ is  the gauge coupling constant of $SU(2)_{L}$ and  
$m_{S}$ is a SUSY particle mass.
The RGE's develop the off-diagonal elements of the slepton mass matrix 
and A-term. 
These terms at the low energy are approximated as 
\begin{eqnarray}
(\bigtriangleup {\bf m}_{\tilde{\sL}}^{2})_{ij}
&\simeq&
-\frac{(6+2a_{0}^2)m_{0}^2}{16\pi^2}({\bf Y_{\nu}^{\dag}Y_{\nu}})_{ij}
\ln
\frac{M_{\sX}}{M_{\sR}}\ ,
\end{eqnarray}
where $M_{\sX}$ is the GUT scale. 
Therefore, off-diagonal elements of ${\bf (Y_{\nu}^{\dag}Y_{\nu}})_{ij}$ are
the crucial quantity to estimate the branching ratio.

As discussed in section 2, 
${\bf (Y_{\nu}^{\dag}Y_{\nu})}_{ij}$ is given by neutrino masses and mixings
at the electroweak scale.
Therefore, we can compare the quantity ${\bf (Y_{\nu}^{\dag}Y_{\nu})}_{ij}$ 
among the cases of  three neutrino mass spectra: 
the degenerate, the inverse-hierarchical and the hierarchical masses. 
In this section, 
we present numerical results in these three cases. 
Here, we use eq.(\ref{rate}) and the vertex functions in Appendix C 
for the calculation of the branching ratio including the RGE's effect.

\subsection{$\mu \rightarrow e+\gamma$}

We present a qualitative discussion
on $({\bf Y}_{\nu}^{\dag}{\bf Y}_{\nu})_{21}$
before predicting  the branching ratio BR$(\mu \rightarrow e+ \gamma)$.   
This is given in terms of neutrino masses and mixings at the electroweak scale
as follows:
\begin{equation}
({\bf Y}_{\nu}^{\dag}{\bf Y}_{\nu})_{21}=
\frac{M_{\sR}}{v_{u}^{2}}
\left[U_{\mu2}U_{e2}^{*}(m_{\nu 2}-m_{\nu 1})+
 U_{\mu3}U_{e3}^{*}(m_{\nu 3}-m_{\nu 1})
\right] \ ,
\label{normal}
\end{equation}
\noindent 
where $v_u\equiv v\sin\b$ with $v=174\G$ is taken  
as an usual notation and  the unitarity condition of the lepton mixing matrix
elements is used.
Taking the three cases of the neutrino mass spectra: the degenerate, 
the inverse-hierarchical  and the normal hierarchical masses,
one obtains the following forms, repectively,
\begin{eqnarray}
({\bf Y}_{\nu}^{\dag}{\bf Y}_{\nu})_{21}
\simeq &&
\frac{M_{\sR}}{\sqrt{2}v_{u}^{2}}\frac{\Delta m^2_{\rm atm}}{2m_\nu}
\left[
\frac{1}{\sqrt{2}}U_{e2}^* \frac{\Delta m_{\odot}^2}{\Delta m^2_{\rm atm}}
+ U_{e3}^*
\right] \ ,          
\quad ({\rm Degenerate}) 
\nonumber \\
\nonumber \\
\simeq  &&
\frac{M_{\sR}}{\sqrt{2}v_{u}^{2}}\sqrt{\Delta m_{\rm atm}^2}
\left[ 
\frac{1}{2\sqrt{2}} U_{e2}^*\frac{\Delta m_{\odot}^2}{\Delta m^2_{\rm atm}} 
- U_{e3}^*
\right] \ , 
\quad ({\rm Inverse }) 
\label{muemass3} \\
\nonumber \\
\simeq  &&
\frac{M_{\sR}}{\sqrt{2}v_{u}^{2}}\sqrt{\Delta m_{\rm atm}^2}
\left[
\frac{1}{\sqrt{2}}U_{e2}^*
\sqrt{\frac{\Delta m_{\odot}^2}{\Delta m^2_{\rm atm}}}
+ U_{e3}^*
\right] \ , 
\quad ({\rm Hierarchy }) \nonumber
\end{eqnarray}
\noindent
where we take the maximal mixing for the atmospheric neutrinos. 
Since  $U_{e2}\simeq 1/\sqrt{2}$ for the bi-maximal mixing matrix, 
the first terms in the square brackets of the right hand sides of  
eqs.(\ref{muemass3}) can be estimated by putting the experimental data.
For the case of the degenerate neutrino masses,
$({\bf Y}_{\nu}^{\dag}{\bf Y}_{\nu})_{21}$ depends on the unknown neutrino mass scale 
$m_{\nu}$.  
As one takes the smaller $m_\nu$, one predicts the larger branching ratio. 
In our calculation, we take $m_{\nu}=0.3 \eV$
\footnote{Recently, a positive observation of the neutrinoless double beta 
decay was reported in \cite{data}, where the degenerate neutrino mass of 
 $m_{\nu}=0.3 \eV$ is a typical one.}
, which is close to the upper 
bound from the neutrinoless double beta decay experiment \cite{Beta},
and also leads to the smallest branching ratio. 

We also note that the degenerate case  gives 
the smallest branching ratio  BR$(\mu \rightarrow e +\gamma)$ among the
three cases as seen in  eqs.(\ref{muemass3}) owing to the scale of
$m_{\nu}$.
It is easy to see the fact that the second terms in eqs.(\ref{muemass3}) 
are dominant as far as  
$U_{e3} ~\>~ 0.01({\rm degenerate}), ~ 0.01({\rm inverse})$ 
and $0.07({\rm hierarchy})$, respectively.
The magnitude and the phase of $U_{e3}$ are important in the comparison between cases of the inverse-hierarchical and the normal hierarchical masses.
In the limit of  $U_{e3}=0$, the predicted branching ratio in the 
case of the normal hierarchical masses is larger than the other
one. However, for $U_{e3}\simeq 0.2$ 
the predicted branching ratios are almost the same in both cases.

\begin{figure}
\vspace{-2.0 cm}
\epsfxsize=9.5 cm
\centerline{\epsfbox{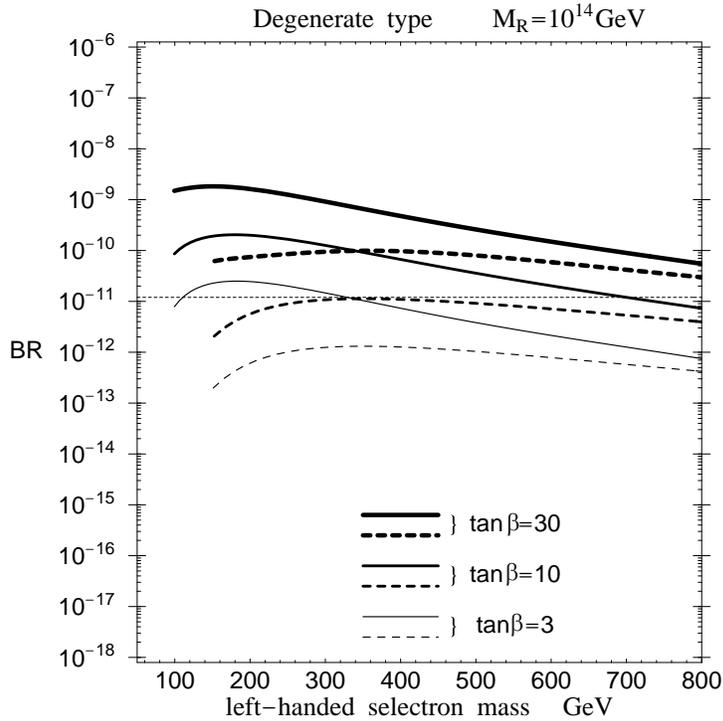}} 
\vspace{-0.3 cm}
\caption{Predicted branching ratio BR$(\mu \rightarrow e+\gamma)$
versus the left-handed selectron mass for $\tan\b=3,\ 10,\ 30$
in the case of the degenerate neutrino masses.
Here $M_\sR=10^{14}\G$ and $U_{e3}=0.2$ are taken.
The solid curves correspond to  $M_2=150 \G$ and the dashed ones to  
$M_2=300 \G$.
A horizontal dotted line denotes the experimental upper bound.}
\label{D1402}
\end{figure}
\begin{figure}
\vspace{-0.8 cm}
\epsfxsize=9.5 cm
\centerline{\epsfbox{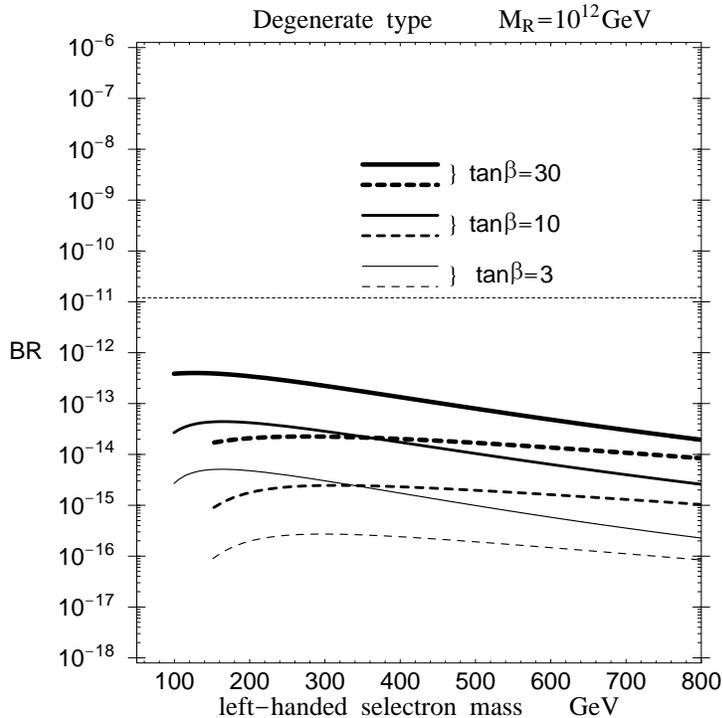}}
\vspace{-0.3 cm}
\caption{Predicted branching ratio BR$(\mu \rightarrow e+\gamma)$
versus the left-handed selectron mass for $\tan\b=3,\ 10,\ 30$
in the case of the degenerate neutrino masses.
Here $M_\sR=10^{12}\G$ and $U_{e3}=0.2$ are taken.
The solid curves correspond to  $M_2=150 \G$ and the dashed ones to  
$M_2=300 \G$.}
\label{D1202}
\end{figure}
\begin{figure}
\vspace{-1.0 cm}
\epsfxsize=9.5 cm
\centerline{\epsfbox{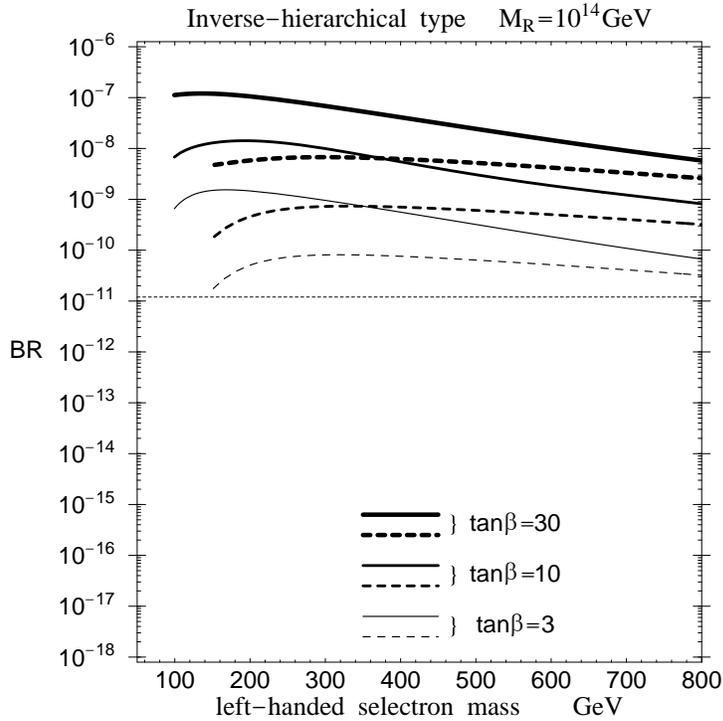}}
\caption{Predicted branching ratio BR$(\mu \rightarrow e+\gamma)$
versus the left-handed selectron mass for $\tan\b=3,\ 10,\ 30$
in the case of the inverse-hierarchical neutrino masses.
Here $M_\sR=10^{14}\G$ and $U_{e3}=0.2$ are taken.
The solid curves correspond to  $M_2=150 \G$ and the dashed ones to  
$M_2=300 \G$.}
\label{I1402}
\end{figure}
\begin{figure}
\vspace{-1.0 cm}
\epsfxsize=9.5 cm
\centerline{\epsfbox{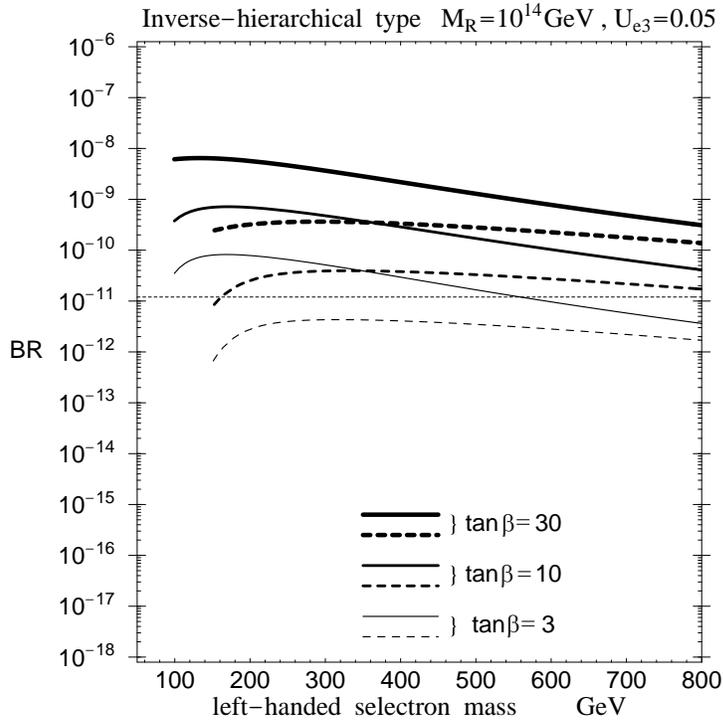}}
\caption{Predicted branching ratio BR$(\mu \rightarrow e+\gamma)$
versus the left-handed selectron mass for $\tan\b=3,\ 10,\ 30$
in the case of the inverse-hierarchical neutrino masses.
Here $M_\sR=10^{14}\G$ and $U_{e3}=0.05$ are taken.
The solid curves correspond to  $M_2=150 \G$ and the dashed ones to  
$M_2=300 \G$.}
\label{I14005}
\end{figure}
\begin{figure}
\vspace{-1.0 cm}
\epsfxsize=9.5 cm
\centerline{\epsfbox{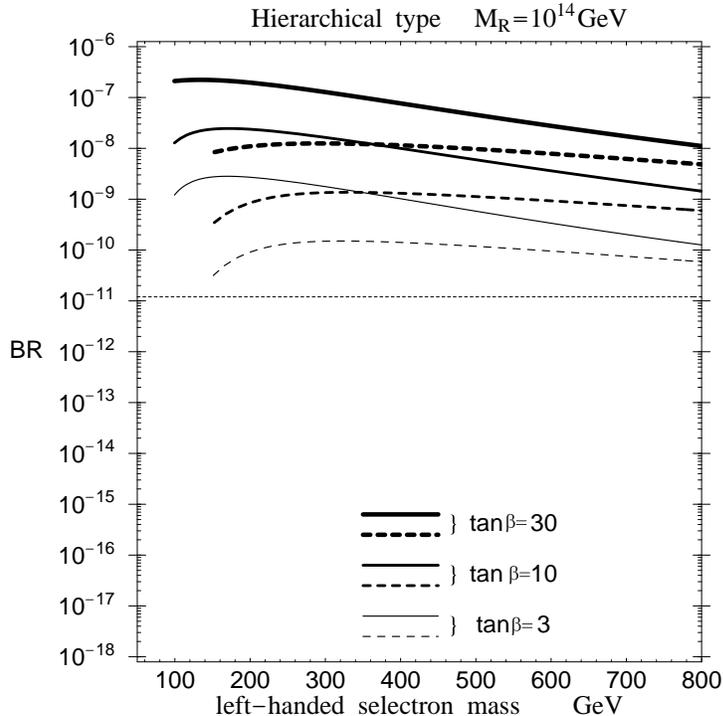}}
\caption{Predicted branching ratio BR$(\mu \rightarrow e+\gamma)$
versus the left-handed selectron mass for $\tan\b=3,\ 10,\ 30$
in the case of the hierarchical neutrino masses.
Here $M_\sR=10^{14}\G$ and $U_{e3}=0.2$ are taken.
The solid curves correspond to  $M_2=150 \G$ and the dashed ones to  
$M_2=300 \G$.}
\label{H1402}
\end{figure}

 At first, we  present numerical results in the case of the degenerate 
neutrino masses assuming ${\bf M_\sR}=M_\sR {\bf 1}$.
The magnitude of $M_\sR$ is constained considerably if we impose the
  $b-\tau$  unfication of Yukawa couplings  \cite{btau}.
In the case of $\tan\beta \leq 30$, the lower bound of
$M_\sR$ is approximately $10^{12}\G$.  We take also $M_\sR \leq 10^{14}\G$,
in order that neutrino Yukawa couplings remain below ${\cal O}(1)$.
 Therefore, we use  $M_\sR = 10^{12},\ 10^{14}\G$ 
 in our following calculation.

We take a universal scalar mass $(m_0)$ for all scalars and
$a_0=0$ as a universal A-term at the GUT scale ($M_\sX=2\times 10^{16}$ GeV).
The branching ratio of  $\mu \rightarrow e+\gamma$ is given 
versus the left-handed 
selectron mass $m_{\tilde e_\sL}$ for each $\tan\b=3,\  10,\  30$
and a fixed wino mass $M_2$ at the electroweak scale.
In fig.\ref{D1402}, the branching ratios
are  shown for  $M_2=150,\  300\ \G$
 in the case of $U_{e3}=0.2$ with $M_R=10^{14} \G$ and $m_{\nu}=0.3$eV,
in which the solid curves correspond to  $M_2=150 \G$
and the dashed ones to  $M_2=300 \G$.
The threshold of the selectron mass is determined by the
recent LEP2 data \cite{LEP} for   $M_2=150 \G$, however, for $M_2=300
\G$,  determined by the  constraint that  the left-handed slepton  should be 
 heavier than the neutralinos.
As the  $\tan\b$ increases, the branching ratio increases
because the decay amplitude from the SUSY diagrams is approximately 
proportional to  $\tan\b$ \cite{LFV1}.
It is found that the branching ratio is  almost larger than the
experimental upper bound in the case of   $M_2=150 \G$.
On the other hand, the predicted values are smaller than the
experimental bound except for $\tan\b=30$  in the case of $M_2=300 \G$.

 Our predictions depend on $M_\sR$ strongly, because the magnitude of
the neutrino Yukawa coupling is determined by  $M_\sR$
as seen in eq.(\ref{YEW}).
If  $M_\sR$ reduces to  $10^{12} \G$,
the branching ratio becomes $10^{4}$ times smaller since it
is proportional to $M_\sR^2$.
The numerical result is shown in fig.\ref{D1202}.
We will examine a model \cite{FTY,O3}, which  
gives the degenerate neutrino masses with 
$U_{e3}\sim 0.05$ in section 4. 

 Next we show results in the case of the inverse-hierarchical neutrino 
masses.  As expected in eq.(\ref{muemass3}), the branching ratio
is much larger than the one in the degenerate case.
 In fig.\ref{I1402},
the branching ratio is shown for  $M_2=150,\  300\ \G$
 in the case of $U_{e3}=0.2$ with $M_\sR=10^{14} \G$.
 In fig.\ref{I14005}, the branching ratio is shown for 
$U_{e3}=0.05$ with  $M_\sR=10^{14} \G$.
The  $M_R$ dependence is the same as the case of 
 the quasi-degenerate neutrino masses.
The predictions almost exceed the experimental bound  as far as
$U_{e3}\geq 0.05$, $\tan\beta\geq 10$ and  $M_\sR\simeq 10^{14} \G$.
This result is based on the assumption  ${\bf M}_{\bf \sR}=M_{\sR}{\bf 1}$,
however, it is not  guaranteed in the case of the inverse-hierarchical 
neutrino masses.
We will examine  a typical  model  \cite{Shafi},  which gives 
${\bf M}_{\bf \sR} \not =M_{\sR}{\bf 1}$ in section 4.

For comparison,
we show the branching ratio in the case of the hierarchical neutrino 
masses in fig.\ref{H1402}. 
It is similar to the case of the inverse-hierarchical neutrino masses.
The branching ratio in the case of the 
degenerate neutrino masses is $10^2$ times smaller than 
the one in the inverse-hierarchical 
and the hierarchical neutrino spectra.

In our numerical analyses we assumed $a_{0}=0$ at the GUT scale
$M_{\sX}$ for simplicity.
Let us comment on the A-term dependence, namely $a_{0} \neq 0$ at $M_{\sX}$.
We estimate the branching ratio for $a_{0}= \pm 1$ at $M_{\sX}$ 
$({\bf A}={\bf Y}a_{0}m_{0})$.
In the degenerate type, the predicted branching ratio is 
1.02($a_{0}=1$), 1.07($a_{0}=-1$) times as large as the one 
in the case of $a_{0}=0$ (tan$\beta=30$, $U_{e3}=0.2$). 
In the inverse-hierarchical type, the predicted branching ratios are
1.56($a_{0}=1$), 1.54($a_{0}=-1$) times as large as the one 
in the case of $a_{0}=0$ (tan$\beta=30$, $U_{e3}=0.2$). 
Therefore the A-term dependence is insignificant in our analyses.

In our calculations,
we use the universality condition at $M_{\sX}$.
We also examine the no-scale condition $m_{0}=0$ at $M_{\sX}$. 
It is found that the predicted branching ratio is $10$ times smaller 
than the one in the case of non-zero universal scalar mass.

\subsection{Non-degeneracy Effect of ${\bf M}_{R}$}

The analyses in the previous section depend on the
  assumption of $M_{\sR1}=M_{\sR2}=M_{\sR3}\equiv M_R$.
In the case of the quasi-degenarate neutrino masses 
in eq.(\ref{eqn:degeneratetype}) 
this complete degeneracy of ${\bf M}_{\sR}$ may be deviated in the 
following magnitude without fine-tuning:
\begin{eqnarray}
\frac{M_{\sR3}^2}{M_{\sR1}^2} \simeq 1 \pm  \frac{\Delta m _{\rm atm}^2}{m_\nu^2}\ ,
\qquad 
\frac{M_{\sR2}^2}{M_{\sR1}^2} \simeq 1 \pm  \frac{\Delta m_{\odot}^2}{m_\nu^2} \ .
\end{eqnarray}
\noindent
Therefore, we  parametrize  $M_{\sR}$ as 
\begin{eqnarray}
{\bf M_{\sR}}
=
M_{\sR}
\left(
\begin{array}{ccc}
1&       0       &0 \\
0 &1+\varepsilon_{2}&0 \\
0 &       0       & 1+\varepsilon_{3}
\end{array}
\right) \ ,
\end{eqnarray}
\noindent
where $\varepsilon_2\simeq  \Delta m_{\odot}^2/2 m_\nu^2$ and 
$\varepsilon_3\simeq  \Delta m _{\rm atm}^2/2 m_\nu^2$.
By using eq.(\ref{yukawa}), we obtain
\begin{equation}
{\bf Y}^{\dagger}_{\nu}{\bf Y}_{\nu}
=
\frac{M_{\sR}}{v^2_{u}}
{\bf U}_{\MNS}
\left(
\begin{array}{ccc}
\sqrt{m_{\nu1}}&   0            & 0\\
        0       &\sqrt{m_{\nu2}}& 0\\
         0      &           0    & \sqrt{m_{\nu3}}
\end{array}
\right)
{\bf K}
\left(
\begin{array}{ccc}
\sqrt{m_{\nu1}}  &    0             & 0\\
      0           & \sqrt{m_{\nu2}} &0 \\
        0         &           0      & \sqrt{m_{\nu3}}
\end{array}
\right)
{\bf U}^{\rm T}_{\MNS} \ ,
\end{equation}
\noindent where
\begin{equation}
{\bf K}\equiv  {\bf R}^{\dagger}
\left(
\begin{array}{ccc}
1  &    0             & 0\\
 0  & 1+\varepsilon_{2}  & 0\\
0   &             0    & 1+\varepsilon_{3}
\end{array}
\right)
{\bf R} \ .
\end{equation}
\noindent 
Then, we have
\begin{eqnarray}
({\bf Y}^{\dagger}_{\nu}{\bf Y}_{\nu})_{21}
=
\frac{M_{\sR}}{v^2_{u}}
\sum^{3}_{i,j}
U_{2 i}U_{1 j}
(K_{ij}\sqrt{m_{\nu i}}\sqrt{m_{\nu j}}) \ ,
\end{eqnarray}
\noindent with
\begin{eqnarray}
K_{ij}
=
\delta_{ij}
+\varepsilon_{2}R_{2i}R_{2j}
+\varepsilon_{3}R_{3i}R_{3j} \ ,
\end{eqnarray}
where we used ${\bf R}^{\rm T}{\bf R=1}$
\footnote{we assume ${\bf R}$ to be real for simplicity. }.
So, we get 
\begin{eqnarray}
({\bf Y}^{\dagger}_{\nu}{\bf Y}_{\nu})_{21}
=
\left.
({\bf Y}^{\dagger}_{\nu}{\bf Y}_{\nu})_{21}
\right |_{\bf M_{\sR \propto 1}}+
\Delta({\bf Y}^{\dagger}_{\nu}{\bf Y}_{\nu})_{21} \ ,
\end{eqnarray}
where the first term is the  $({\bf Y}^{\dagger}_{\nu}{\bf Y}_{\nu})_{21}$ element
in eq.(\ref{normal}), which corresponds to the  ${\bf M_{\sR \propto 1}}$,
while the second term stands for the deviation from it as follows:
\begin{equation}
\Delta({\bf Y}^{\dagger}_{\nu}{\bf Y}_{\nu})_{21}=
\frac{M_{\sR}}{v^2_{u}}
\sum^{3}_{i,j}
U_{2 i}U_{1 j}
\sqrt{m_{\nu i}}\sqrt{m_{\nu j}}
(\varepsilon_{2}R_{2i}R_{2j}
+\varepsilon_{3}R_{3i}R_{3j}) \  .
\end{equation}
In order to estimate the second term, we use
$\varepsilon_2=0.0001$ and $\varepsilon_3=0.01$ taking account of
 $\varepsilon_2\simeq \Delta m_{\odot}^2/2 m_\nu^2$ and 
$\varepsilon_3\simeq  \Delta m _{\rm atm}^2/2 m_\nu^2$, where
$m_\nu=0.3 \eV$ is put.
Since $m_{\nu i}\simeq m_{\nu j}$  and $R_{ij} \leq 1$,
we get 
\begin{eqnarray}
\Delta({\bf Y}^{\dagger}_{\nu}{\bf Y}_{\nu})_{21}
&\sim&
\frac{M_{\sR}}{v^2_{u}}
\sum^{3}_{i,j}
U_{2 i}U_{1 j}
m_{\nu}
\varepsilon_{3}R_{3i}R_{3j}
\nonumber\\
&\leq&
\frac{M_{\sR}}{v^2_{u}}
\frac{1}{2\sqrt{2}}
m_{\nu}
\varepsilon_{3}
\sim 3.5 \times 10^{-3} \ .
\end{eqnarray}
Taking this maximal value, we can estimate the branching ratio 
as follows:
\begin{eqnarray}
\frac{BR({\rm non\mbox{-}degenerate}\ M_{\sR})}{BR({\rm degenerate} \ M_{\sR})}
\leq
\left(\frac{2.6 + 3.5}{2.6} \right)^2 \simeq  5.5 \ .
\end{eqnarray}
Therefore, the enhancement due to the second term is at most 
factor 5.
This conclusion does not depend on the specific form of ${\bf R}$

Consider the case of the  inverse-hierarchical type of neutrino masses.
We take $\varepsilon_{2}\sim 0.01$ 
with the similar argument of the quasi-degenerate type neutrino masses,
because 
$m_{\nu 1}$ and $m_{\nu 2}$ are almost degenerate and 
$\varepsilon_2\simeq  \Delta m _{\odot}^2/2 \Delta m _{\rm atm}^2$ 
in this case.
Then, we get  
\begin{eqnarray}
\Delta({\bf Y}^{\dagger}_{\nu}{\bf Y}_{\nu})_{21}
&=&
\frac{M_{\sR}}{v^2_{u}}
\sum^{2}_{i,j}
U_{2 i}U_{1 j}
m_{\nu1}
\varepsilon_{2}R_{2i}R_{2j}
\nonumber\\
&\leq&
\frac{M_{\sR}}{v^2_{u}}
\frac{1}{2\sqrt{2}}
m_{\nu1}
\varepsilon_{2}
\sim  0.063 \times 10^{-2} \ ,
\end{eqnarray}
where we assume $\varepsilon_{2} \geq \varepsilon_{3}$ and  use
$m_{\nu3}\simeq 0$, $m_{\nu1} \simeq  m_{\nu2} \simeq 0.054$eV  
and $R_{ij}\leq 1$.
Taking  the maximal value, we get 
\begin{eqnarray}
\frac{BR({\rm non\mbox{-}degenerate} \ M_{\sR})}{BR({\rm degenerate} \ M_{\sR})}
\leq 
\left(
\frac{2.7 + 0.063}{2.7}
\right)^2
\simeq 1.04 \ .
\end{eqnarray}
Thus, the effect of the $\Delta({\bf Y}^{\dagger}_{\nu}{\bf Y}_{\nu})_{21}$ is very small
in the case of the inverse hierarchical neutrino masses.

These discussions in this subsection are  also available qualitatively
for the $\tau \rightarrow \mu+\gamma$ process.


\begin{figure}
\vspace{-0.35 cm}
\epsfxsize=9.5 cm
\centerline{\epsfbox{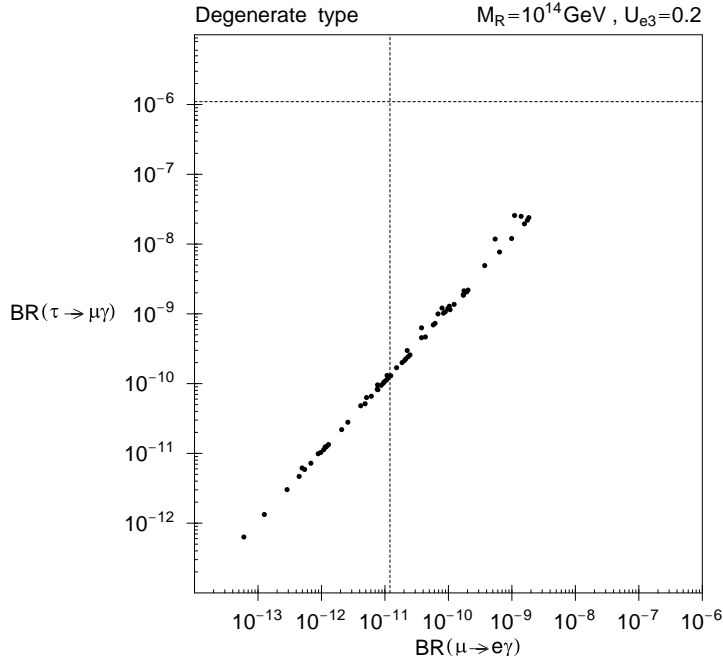}}
\caption{Predicted branching ratio BR$(\tau \rightarrow \mu+\gamma)$
versus BR$(\mu \rightarrow e+\gamma)$ for $\tan\b=3,\ 10,\ 30$
in the case of the degenerate neutrino masses.
Here $M_\sR=10^{14}\G$, $m_{\nu}=0.3$eV and $U_{e3}=0.2$ are taken
and the left-handed selectron mass is taken as same as in figs.2-4.}
\label{DTM1402}
\end{figure}
\begin{figure}
\vspace{-0.25 cm}
\epsfxsize=9.5 cm
\centerline{\epsfbox{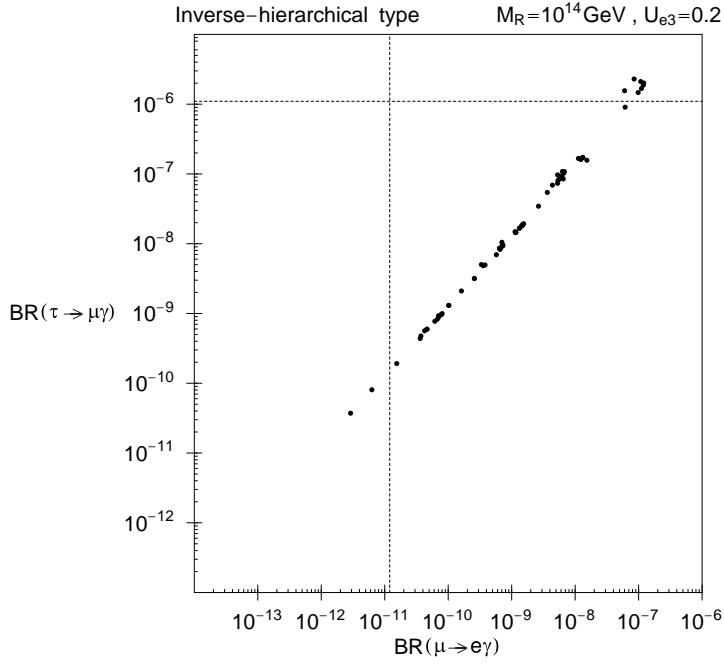}}
\caption{Predicted branching ratio BR$(\tau \rightarrow \mu+\gamma)$
versus  BR$(\mu \rightarrow e+\gamma)$ for $\tan\b=3,\ 10,\ 30$
in the case of the inverse-hierarchical neutrino masses.
Here $M_\sR=10^{14}\G$ and $U_{e3}=0.2$ are taken
and the left-handed selectron mass is taken as same as in figs.2-4.}
\label{ITM1402}
\end{figure}
\subsection{$\tau \rightarrow \mu+\gamma$}
Let us study  the  $\tau \rightarrow \mu +\gamma$ process.
In this case, we should discuss
\begin{equation}
({\bf Y}_{\nu}^{\dag}{\bf Y}_{\nu})_{32}=
\frac{M_{\sR}}{v_{u}^{2}}
\left[
U_{\tau 2}U_{\mu 2}^{*}(m_{\nu2}-m_{\nu1})
+U_{\tau 3}U_{\mu 3}^{*}(m_{\nu3}-m_{\nu1})
\right] \ .
\label{eqn:taumuYY} 
\end{equation}
It should be stressed that it is independent of $U_{e3}$ 
in contrast to
$({\bf Y}_{\nu}^{\dagger}{\bf Y}_{\nu})_{21}$.
Therefore we can determine the following form of  
$({\bf Y}_{\nu}^{\dagger}{\bf Y}_{\nu})_{32}$ at the electroweak scale by
using the bi-maximal mixing matrix:

\begin{eqnarray}
({\bf Y}_{\nu}^{\dag}{\bf Y}_{\nu})_{32}
\simeq  &&
\frac{M_{\sR}}{v_{u}^{2}}
\left[ 
-\frac{1}{4}\frac{\Delta m_{\odot}^2}{2m_{\nu}}
+\frac{1}{2}\frac{\Delta m^2_{\rm atm}}{2m_{\nu}}
\right] \ 
\simeq
\frac{M_{\sR}}{4v_{u}^{2}}
\frac{\Delta m^2_{\rm atm}}{m_{\nu}}
\quad,\  ({\rm Degenerate })  
\nonumber\\
\nonumber\\
\simeq  &&
\frac{M_{\sR}}{v_{u}^{2}}
\left[
\frac{1}{8}
\frac{\Delta m_{\odot}^2}
{\sqrt{\Delta m^2_{\rm atm}}}
-\frac{1}{2}\sqrt{\Delta m^2_{\rm atm}}
\right] \ 
\simeq
-\frac{M_{\sR}}{2v_{u}^{2}}
\sqrt{\Delta m^2_{\rm atm}}
\quad,\ ({\rm Inverse }) 
\nonumber\\
\label{mass3} 
\\
\simeq &&
\frac{M_{\sR}}{v_{u}^{2}}
\left[
-\frac{1}{4}\sqrt{\Delta m_{\odot}^2}
+\frac{1}{2}\sqrt{\Delta m^2_{\rm atm}}
\right] \ 
\simeq
\frac{M_{\sR}}{2v_{u}^{2}}
\sqrt{\Delta m^2_{\rm atm}}
\quad.\ ({\rm Hierarchy})
\nonumber 
\end{eqnarray}
\noindent
We see that the case of the inverse-hierarchical masses and the hierarchical 
masses are almost the same as seen in eqs.(\ref{mass3}).  

Let us present numerical results of
 BR$(\tau \rightarrow \mu+\gamma)$ \cite{TauMuGamma} 
versus BR$(\mu \rightarrow e+\gamma)$ \cite{MuEGamma} in the case of the 
 degenerate neutrino mass, 
in which  $\tan\b=3,\  10,\  30$ are taken.
In fig.\ref{DTM1402}, 
the branching ratio is plotted  
for  $M_2=150,\ 300\ \G$ for $U_{e3}=0.2$ with $M_\sR=10^{14} \G$.
Dotted lines are the experimantal upper bounds for BR$(\tau \rightarrow
\mu+\gamma)$ and BR$(\mu \rightarrow e+\gamma)$,  respectively. 
The dependence of $\tan\b$ is the same as the case of 
$\mu \rightarrow e+\gamma$.
It is found that the branching ratio is completely smaller than the
experimental upper bound in the case of $\tau \rightarrow \mu+\gamma$ in 
contrast with the case of $\mu \rightarrow e+\gamma$.

 Next we show the  results in the case of the inverse-hierarchical neutrino 
masses.  As expected in eqs.(\ref{mass3}), the branching ratio
is much larger than the one in the degenerate case.
 In fig.\ref{ITM1402},
the branching ratio is shown for  $M_2=150,\  300\ \G$
 in the case of $U_{e3}=0.2$ with $M_\sR=10^{14} \G$.
In conclusion, 
the predicted branching ratio is larger than the one in the case 
of the degenerate neutrino mass, and it is almost smaller than the
experimental upper bound for $\tau \rightarrow \mu+\gamma$ in 
contrast with $\mu \rightarrow e+\gamma$.
The constraint of BR$(\mu \rightarrow e+\gamma)$ is always
severer than the one in the case of BR$(\tau \rightarrow \mu+\gamma)$.

\section{Typical models and numerical analyses}

\subsection{$S_{3\sL}\times S_{3\sR}$ flavor symmetry model\ -\ Degenarate type}

In this section we examine the  neutrino model proposed by
Fukugida, Tanimoto and Yanagida \cite{FTY},  which derives 
the quasi-degenerate masses, $m_{\nu 1} \sim m_{\nu 2} \sim m_{\nu3}$.
This model is based on the $S_{3\sL} \times S_{3\sR}$ flavor symmetry 
\cite{Demo}.
Taking ${\bf M_\sR}=M_\sR {\bf 1}$, the neutrino Yukawa coupling is given as
follows:
\begin{eqnarray}
{\bf Y}_{\nu}
=Y_{\nu 0} \left [
\left(
\begin{array}{ccc}
1              & 0            & 0 \\
0              & 1            & 0 \\
0              & 0            & 1
\end{array}
\right)
+
\left(
\begin{array}{ccc}
 0     & 0              & 0 \\
0                   & \epsilon_{\nu} & 0 \\
0                   & 0              & \delta_{\nu}
\end{array}
\right) \right ] \ ,
\end{eqnarray}
where we take the diagonal basis for the neutrino sector.
The first matrix is the $S_{3\sL}$ invariant one, and the second one
is the  symmetry breaking term.
The parameters $Y_{\nu 0}, \epsilon_{\nu}$ and $\delta_{\nu}$ are 
constrained by the experimental values of 
$\Delta m_{\rm atm}^2$ and $\Delta m_{\odot}^2$. 
Therefore, the flavor mixings come from the charged lepton Yukawa couplings.

The charged lepton Yukawa coupling is given by the symmetry breaking
parameters $\epsilon_{l}, \delta_{l}$ as follows:
\begin{eqnarray}
{\bf Y}_{e}
=Y_{e0} \left [
\left(
\begin{array}{ccc}
1 & 1 & 1 \\
1 & 1 & 1 \\
1 & 1 & 1
\end{array}
\right)
+
\left(
\begin{array}{ccc}
-\epsilon_{l}     & 0             & 0 \\
0                 & +\epsilon_{l}  & 0 \\
0                 & 0             & +\delta_{l}
\end{array}
\right) \right ] \ .
\end{eqnarray}
\begin{figure}
\epsfxsize=9.5 cm
\vspace{-1.0 cm}
\centerline{\epsfbox{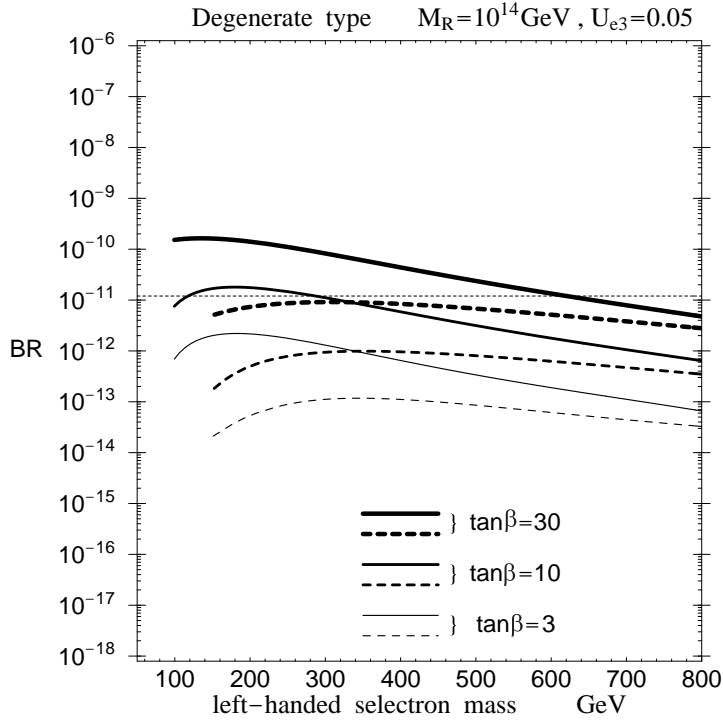}}

\caption{Predicted branching ratio BR$(\mu \rightarrow e+\gamma)$
versus the left-handed selectron mass for $\tan\b=3,\ 10,\ 30$
in the case of the $S_{3L}\times S_{3R}$ flavor symmetry model.
Here $M_\sR=10^{14}\G$ are taken.
The solid curves correspond to  $M_2=150 \G$ and the dashed ones to  
$M_2=300 \G$.}
\label{D14005}
\end{figure}
Since $Y_{e0}, \epsilon_{l}$ and  $\delta_{l}$ are fixed by 
the charged lepton masses, 
one gets the lepton mixing matrix elements as follows :
\begin{eqnarray}
{\bf U}_{\MNS}
\simeq
\left(
\begin{array}{ccc}
1/\sqrt{2}   & -1/\sqrt{2} & \sqrt{2/3}\sqrt{m_{e}/m_{\mu}} \\
 1/\sqrt{6}  & 1/\sqrt{6}  & -2/\sqrt{6} \\
 1/\sqrt{3} & 1/\sqrt{3} & 1/\sqrt{3}
\end{array}
\label{eqn:SLSRMNS}
\right)~ \ .
\end{eqnarray}
As a result, we see 
$U_{e3}=\sqrt{2/3}\sqrt{m_{e}/m_{\mu}}\sim 0.05$ from eq.(\ref{eqn:SLSRMNS}).

We estimated the branching ratio of the processes 
$\mu \rightarrow e+\gamma$ 
and $\tau \rightarrow \mu +\gamma$ 
by using $U_{e3}= 0.05$.
We show the branching ratio   for $M_{2}=150$ GeV and $300$GeV taking
tan$\beta=3,10,30$ in fig.\ref{D14005}.
Because of the smallness of $U_{e3}$,
we see that BR$(\mu \rightarrow e +\gamma)$ 
is smaller than the experimental upper bound except for tan$\beta=30$,
and $M_{2}=150$GeV.

We have also estimated the branching ratio 
BR($\tau \rightarrow \mu +\gamma)$ for tan$\beta=30$,
which is much smaller than the experimental bound BR($\tau \rightarrow \mu +\gamma)< 1.1\times 10^{-6}$. 
Thus, the $\mu \rightarrow e +\gamma$ process provides the severe
constraint compared with the $\tau \rightarrow \mu +\gamma$ 
in the present experimental situation.

\subsection{The Shafi-Tavartkiladze model\ -\ Inverse-hierarchical type}
The typical model of the inverse-hierarchical neutrino masses
is the Zee model \cite{Zee},
in which the right-handed neutrinos do not exist.
However, one can also consider a Yukawa texture  which leads to the
inverse-hierarchical masses through the see-saw mechanism, 
namely the Shafi-Tavartkiladze model \cite{Shafi}.  

Shafi and Tavartkiladze utilize the anomalous $U(1)$ flavor symmetry 
\cite{U1}. 
In this model, due to the Froggatt-Nielsen mechanism \cite{FrNi},
one of the Yukawa interaction term in the effective theory is given as 
\begin{eqnarray}
 e^{c}_{{\sR}i}L_{j}H_{d}
\left(
\frac{S}{M_{\rm pl}}
\right)^{m_{ij}} \ ,
\label{YukawaInt}
\end{eqnarray}
where $e^{c}_{\sR{i}}$ and $L_{j}$ are the right-handed charged lepton and
the left-handed lepton doublet, respectively, 
$H_{d}$ is Higgs doublet, and 
$S$ is singlet field.
The effective Yukawa couplings are given in terms of 
\begin{eqnarray}
 \lambda\equiv\frac{\langle S\rangle}{M_{\rm pl}}\simeq 0.2\ .
\end{eqnarray}

\begin{figure}
\epsfxsize=9.5 cm
\vspace{-1.0 cm}
\centerline{\epsfbox{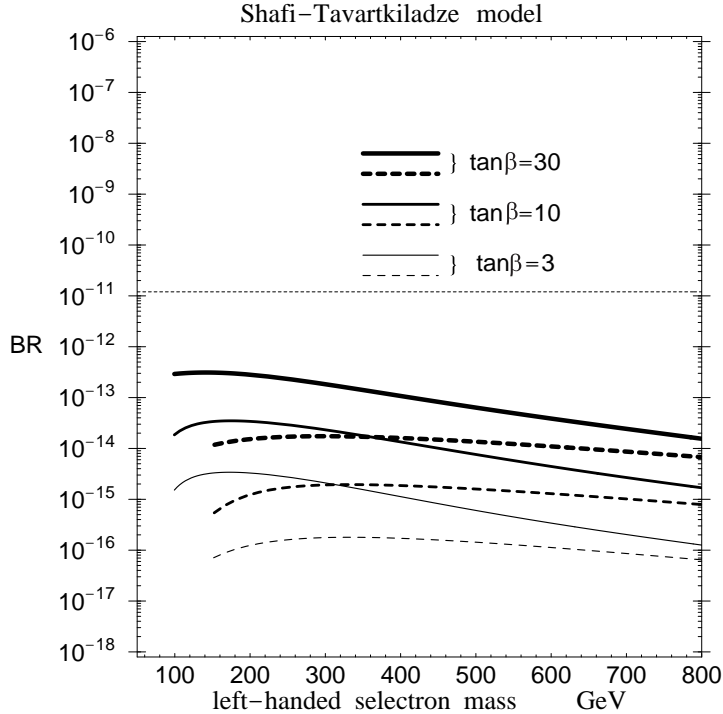}}
\caption{Predicted branching ratio BR$(\mu \rightarrow e+\gamma)$
versus the left-handed selectron mass for $\tan\b=3,\ 10,\ 30$
in the case of the Shafi-Tavartkiladze model\cite{Shafi}.
The solid curves correspond to  $M_2=150 \G$ and the dashed ones to  
$M_2=300 \G$.}
\label{shafi}
\end{figure}

The neutrino mass matrix is given in Appendix D.
Fixing the $U(1)$ flavor charges $k$, $n$, $k^{'}$ as $k$=0, $n$=2, $k^{'}$=2,
which is consistent with neutrino mass data,
the Yukawa coupling is given as 
\begin{eqnarray}
 {\bf Y}_{\nu}&=&
\left(
\begin{array}{ccc}
 \lambda^{4}& \lambda^{2}& \lambda^{2}\\
 1 & 0 & 0
\end{array}
\right)\ ,
\label{ShafiYukawa}
\end{eqnarray}
and the right-handed neutrino Majorana mass matrix is given as 
\begin{eqnarray}
 {\bf M}_{\bf \sR}&=&
M_{\sR}
\left(
\begin{array}{cc}
 \lambda^{4}& 1\\
 1& 0 
\end{array}
\right)\ .
\label{ShafiRight}
\end{eqnarray}
It is remarked that right-handed neutrinos contain only two generations in
this model. 
In eq.(\ref{ShafiYukawa}), 
components 2-2 and 2-3 must be zero for the sake of 
holomorphy of superpotential, it is called SUSY zero. 
The neutrino mass matrix is given by the see-saw mechanism as 
\begin{eqnarray}
 {\bf m}_{\nu}=
{\bf Y}_{\nu}^{\rm T}{\bf M}_{\bf\sR}^{-1}{\bf Y}_{\nu}v_{u}^{2}=
\frac{\lambda^{2}v_{u}^{2}}{M_{\sR}}
\left(
\begin{array}{ccc}
\lambda^{2} & 1& 1\\
1& 0&0\\
1&0&0
\end{array}
\right)\ ,
\end{eqnarray}
where the  order one coefficient in front of each entry is neglected.
This mass matrix gives the inverse-hierarchical neutrino masses.
The ${\bf Y}_{\nu}^{\dag}{\bf Y}_{\nu}$ is given as 
\begin{eqnarray}
 {\bf Y}_{\nu}^{\dag}{\bf Y}_{\nu}=
\left(
\begin{array}{ccc}
 1+\lambda^{8}&\lambda^{6}&\lambda^{6}\\
 \lambda^{6}&\lambda^{4}&\lambda^{4}\\
 \lambda^{6}&\lambda^{4}&\lambda^{4}\\ 
\end{array}
\right)\ .
\end{eqnarray}
It is noticed that the component $( {\bf Y}_{\nu}^{\dag}{\bf Y}_{\nu})_{21}$ 
is suppressed as 
\begin{eqnarray}
 ( {\bf Y}_{\nu}^{\dag}{\bf Y}_{\nu})_{21}\sim\l^{6}\sim{\cal O}(10^{-5})\ .
\end{eqnarray}
Then, 
we expect that the branching ratio of $\mu\rightarrow e+\gamma$ in this
model is much smaller than the one in the case of 
$({\bf M}_{\bf\sR})_{3\times3}=M_{\sR}({\bf 1})_{3\times3}$ in section 3. 

In fig.\ref{shafi}, the branching ratio is shown for  $M_2=150,\  300\
\G$.
The predictions are given by taking $\lambda=0.2$ and
all of order one  coefficients in the Yukawa couplings  are fixed to be one.
The predicted value is much smaller than the one in the inverse-hierarchical
case discussed in the section 3. 
Because $({\bf Y}_{\nu}^{\dag}{\bf Y}_{\nu})_{21}$ is proportional to $\l^6$,
the smallness of the branching ratio is understandable.  

\section{Summary and Discussions}
 We have investigated the lepton flavor violating processes 
 $\mu\rightarrow e+\gamma$ and $\tau\rightarrow \mu+\gamma$, 
in the framework of the MSSM with the right-handed neutrinos. 
Even if we impose the universality condition for the soft scalar masses 
and A-terms at the GUT scale, 
off-diagonal elements of the left-handed slepton mass matrix are generated
 through the RGE's running effects from the GUT scale to the right-handed
 neutrino mass scale $M_{\sR}$.
We have taken the LMA-MSW solution for the neutrino masses and mixings.

The branching ratios of $\mu\rightarrow e+\gamma$ and
$\tau\rightarrow\mu+\gamma$ processes are proportional to ${\bf
|(Y_{\nu}^{\dag}Y_{\nu})}_{ij}|^{2}$.
Since ${\bf (Y_{\nu}^{\dag}Y_{\nu})}_{ij}$ depends on the mass spectrum of
neutrinos, 
we can compare the branching ratio of three cases of neutrino mass
spectra: the degenerate, the inverse-hierarchical and the hierarchical case.  

First, 
we have studied the three types in the case of $\mu\rightarrow e+\gamma$, 
in which we take ${\bf M}_{\bf\sR}=M_{\sR}{\bf 1}$. 
For the case of the degenerate neutrino masses, the branching ratio depends 
on the unknown neutrino mass $m_{\nu}$.  
We have taken $m_{\nu}=0.3$eV, which gives us the largest  branching ratio.  
It is  emphasized that  the magnitude of  $U_{e3}$ is one of  important
ingredients to predict BR($\mu \rightarrow e +\gamma $).  
The branching ratio of the inverse-hierarchical case almost exceeds the
experimental upper bound and is much larger than the degenerate case for
$M_2=150 \G$ and $M_2=300 \G$.
In general, 
we expect the relation BR(degenerate)$\ll$BR(inverse-hierarchical)$<$
BR(hierarchical). 
The effect of the deviation from  ${\bf M}_{\bf\sR}=M_{\sR}{\bf 1}$ has been 
estimated. The enhancement of the branching ratios are at most
factor five in the case of the quasi-degenerate neutrino mass spectrum.
 
Second, 
we have studied the three cases in $\tau\rightarrow \mu+\gamma$.
It is noticed that branching ratio is independent of $U_{e3}$ in
contrast to the case of $\mu\rightarrow e+\gamma$. 
For the degenerate neutrino masses, 
the branching ratio is completely smaller than the
experimental upper bound.
For the inverse-hierarchical neutrino masses, 
the branching ratio is almost smaller than the experimental bound. 
The constraint of BR$(\mu \rightarrow e+\gamma)$ is always 
severer than the one in the case of BR$(\tau \rightarrow \mu+\gamma)$.

Finally, 
we have investigated the branching ratio of $\mu\rightarrow e+\gamma$ in the
typical models of the degenetate and the inverse-hierarchical cases. 
Since the $S_{3\sL}\times S_{3\sR}$ model, 
which is a typical one of the degenerate case, 
predicts $U_{e3}\simeq0.05$, the  branching ratio is much smaller
than the case of $U_{e3}\simeq0.2$.  
The Shafi-Tavartkiladze model, 
which is a typical one of the inverse-hierarchical case, predicts 
the very small branching ratio.
Thus, the models can be tested by the  $\mu\rightarrow e+\gamma$ process.
 
The branching ratio of $\mu\rightarrow e+\gamma$ and $\tau\rightarrow
\mu+\gamma$ will be improved to the lebel $10^{-14}$ in the PSI and
$10^{-(7-8)}$ in the B factories in KEK and SLAC, respectively.
Therefore, 
future experiments can probe the framework for the neutrino masses.

\vskip 1.0 cm
{\Large\bf Acknowledgements}\\ 
 We would like to thank Drs. J. Sato, T. Kobayashi, T. Goto and
 H. Nakano for useful discussions.
 We also thank the organizers and participants of Summer 
Institute 2001 held at Yamanashi, Japan for helpful discussions. 
 This research is  supported by the Grant-in-Aid for Science Research,
 Ministry of Education, Science and Culture, Japan(No.10640274, No.12047220). 

\newpage

\newpage

\appendix

\section{Yukawa Matrix}
The Yukawa matrix is determined in general as follows \cite{Casas}.
The left-handed neutrino mass matrix is given as 
\begin{eqnarray}
 {\bf m_{\n}}=
\left({\bf Y_\n} v_u\right)^{\rm T}{\bf M}_{\bf \sR}^{-1}
\left({\bf Y_\n} v_u\right)\ ,
\label{eqn:mn1}
\end{eqnarray}
via the see-saw mechanism, where $v_u$ is the vacuum
expectation value (VEV) of Higgs $H_u$. 
One can always take the diagonal form of the right-handed Majorana neutrino 
mass matrix ${\bf M_\sR}={\bf M_\sR^{\diag}}$.  
The neutrino mass matrix ${\bf m_\n}$ is diagonalized by a single unitary
 matrix
\begin{eqnarray}
{\bf m_\n^{\diag}}
&\equiv&
{\bf U}_{\MNS}^{\rm T}{\bf m_\n} {\bf U_{\MNS}}\  ,\label{eqn:mndiag1}
\end{eqnarray}
where ${\bf U_{\MNS}}$ is the MNS matrix. 
In eqs.(\ref{eqn:mn1}) and (\ref{eqn:mndiag1}), 
one can divide ${\bf M_\sR^{\diag}}$ into square roots
\begin{eqnarray}
{\bf m_{\n}^{\diag}}
&=&
{\bf U}_{\MNS}^{\rm T}{\bf Y}_\n^{\rm T}
\left(\bf {\bf M_\sR^{\diag}}\right)^{-1}
{\bf Y_\n U_{\MNS}}v_u^2
\nonumber\\
&=&
{\bf U}_{\MNS}^{\rm T}{\bf Y}_\n^{\rm T}
\sqrt{\left({\bf M_\sR^{\diag}}\right)^{-1}}
\sqrt{\left({\bf M_\sR^{\diag}}\right)^{-1}}{\bf Y_\n}
 {\bf U_{\MNS}}v_u^2 \ .
\label{eqn:mndiag2}
\end{eqnarray}
Multiplying the inverse square root of the matrix ${\bf m}_\n^{\diag}$
 from both right and left hand sides of eq.(\ref{eqn:mndiag2}),
 one  gets the following form 
\begin{eqnarray}
{\bf 1}&=&\sqrt{\left({\bf m_\n^{\diag}}\right)^{-1}}
{\bf U}_{\MNS}^{\rm T}{\bf Y}_\n^{\rm T}
\sqrt{\left({\bf M_\sR^{\diag}}\right)^{-1}}v_u^2 
\sqrt{\left({\bf M_\sR^{\diag}}\right)^{-1}}{\bf Y_\n U_{\MNS}}
\sqrt{\left({\bf m_\n^{\diag}}\right)^{-1}}
\nonumber\\
&\equiv&
{\bf R}^{\rm T}{\bf R}\ ,
\end{eqnarray}
where one  has defined the following complex orthogonal 3$\times$3
matrix
\begin{eqnarray}
{\bf R}
\equiv
v_u\sqrt{\left({\bf M_\sR^{\diag}}\right)^{-1}}{\bf Y_\n U_{\MNS}}
\sqrt{\left({\bf m_\n^{\diag}}\right)^{-1}},\label{eqn:R}
\end{eqnarray}
and ${\bf R}$ depends on models. 
Therefore,  one can write the neutrino Yukawa coupling as 
\begin{eqnarray}
{\bf Y_\n}
=
\frac{1}{v_u}
\sqrt{{\bf M_\sR^{\diag}}}\ {\bf R}\ \sqrt{{\bf m_\n^{\diag}}} \ 
{\bf U}^{\rm T}_{\MNS} \ ,
\end{eqnarray}
or explicitly
\begin{equation}
{\bf Y_\nu}
= \frac{1}{v_u}  \left (\matrix{\sqrt{M_{\sR 1}}& 0 & 0\cr
  0 & \sqrt{M_{\sR 2}}  & 0 \cr  0 & 0 & \sqrt{M_{\sR 3}} \cr  } \right) 
  {\bf R}
  \left (\matrix{\sqrt{m_{\n 1}}& 0 & 0\cr
  0 & \sqrt{m_{\n 2}}  & 0 \cr  0 & 0 & \sqrt{m_{\n 3}} \cr  } \right)  
 {\bf U}^{\rm T}_{\MNS}\ .
\end{equation}

\section{RGEs}
\subsection{From $M_{\rm\sX}$ to $M_{\rm\sR}$}

\begin{eqnarray}
\mu \frac{d}{d \mu} g^{2}_{i}
&=& \frac{1}{8\pi^{2}}b_{i}g^{4}_{i},
~~~(b_1,b_2,b_3)=(\frac{33}{5},1,-3) \ ,
\nonumber\\
\mu \frac{d}{d \mu} M_{i}
&=& \frac{b_{i}}{2\pi}\alpha_{i}M_{i}\ ,~~~\alpha_{i}=\frac{g^{2}_{i}}{4\pi}
~~~(i=1,2,3) \ ,
\nonumber\\
\mu \frac{d}{d \mu} {\bf Y}_{e}^{ij}&=& \frac{1}{16 \pi^2}
\left[
\left \{
-\frac{9}{5}g_1^2 -3 g_2^2 + 3 {\rm Tr}({\bf Y}_d {\bf Y}_d^{\dagger})
+ {\rm Tr}({\bf Y}_{e} {\bf Y}_{e}^{\dagger})
\right \} {\bf Y}_{e}^{ij} \right.
\nonumber \\
&&\left.
+3({\bf Y}_{e} {\bf Y}_{e}^{\dagger} {\bf Y}_{e})^{ij} 
+ ({\bf Y}_{e} {\bf Y}_{\nu}^{\dagger} {\bf Y}_{\nu})^{ij}
\right],
\nonumber\\
\mu \frac{d}{d \mu} {\bf Y}_{\nu}^{ij}&=& \frac{1}{16 \pi^2}
\left[
\left \{
-\frac{3}{5} g_1^2 -3 g_2^2 +3 {\rm Tr}({\bf Y}_{u} {\bf Y}_{u}^{\dagger})
+{\rm Tr}({\bf Y}_{\nu} {\bf Y}_{\nu}^{\dagger})
\right \} {\bf Y}_{\nu}^{ij}
\right.
\nonumber \\
&&\left.
+3({\bf Y}_{\nu} {\bf Y}_{\nu}^{\dagger} {\bf Y}_{\nu})^{ij}
+ ({\bf Y}_{\nu}{\bf Y}_{e}^{\dagger} {\bf Y}_{e})^{ij}
\right ],
\nonumber\\
\mu \frac{d}{d \mu} {\bf Y}_u^{ij}&=& \frac{1}{16 \pi^2}
\left[
\left \{
-\frac{13}{15}g_1^2 -3 g_2^2 -\frac{16}{3}g_3^2+ 3 {\rm Tr}({\bf Y}_u {\bf Y}_u^{\dagger})
+ {\rm Tr}({\bf Y}_{\nu} {\bf Y}_{\nu}^{\dagger})
\right \} {\bf Y}_u^{ij}
\right.
\nonumber \\
&&\left.
+3({\bf Y}_u {\bf Y}_u^{\dagger} {\bf Y}_u)^{ij} 
+ ({\bf Y}_u {\bf Y}_{d}^{\dagger} {\bf Y}_{d})^{ij}
\right ],
\nonumber\\
\mu \frac{d}{d \mu} {\bf Y}_d^{ij}&=& \frac{1}{16 \pi^2}
\left[
\left \{
-\frac{7}{15}g_1^2 -3 g_2^2 -\frac{16}{3}g_3^2+ 3 {\rm Tr}({\bf Y}_d {\bf Y}_d^{\dagger})
+ {\rm Tr}({\bf Y}_{e} {\bf Y}_{e}^{\dagger})
\right \} {\bf Y}_d^{ij}
\right.
\nonumber \\
&&\left.
+3({\bf Y}_d {\bf Y}_d^{\dagger} {\bf Y}_d)^{ij} 
+ ({\bf Y}_d {\bf Y}_{u}^{\dagger} {\bf Y}_{u})^{ij}
\right ],
\nonumber 
\end{eqnarray}
\begin{eqnarray}
\mu \frac{d}{d \mu} ({\bf m}^2_{\tilde \sL})_i^j
&=&
\frac{1}{16 \pi^2} \left [
\left({\bf m}^2_{\tilde \sL} {\bf Y}_{e}^{\dagger} {\bf Y}_{e}
+{\bf Y}_{e}^{\dagger} {\bf Y}_{e} {\bf m}^2_{\tilde \sL} \right)_i^j
+\left({\bf m}^2_{\tilde \sL} {\bf Y}_{\nu}^{\dagger} {\bf Y}_{\nu}
+{\bf Y}_{\nu}^{\dagger} {\bf Y}_{\nu} {\bf m}^2_{\tilde \sL} \right)_i^j
\right.
\nonumber \\
&&+2 \left( {\bf Y}_{e}^{\dagger} {\bf m}^2_{\tilde e} {\bf Y}_{e}
+m^2_{H_d}{\bf Y}_{e}^{\dagger}{\bf Y}_{e}
+{\bf A}_{e}^{\dagger} {\bf A}_{e} \right)_i^j
\nonumber \\
&&+2 \left( {\bf Y}_{\nu}^{\dagger} m^2_{\tilde \nu} {\bf Y}_{\nu}
+{m}^2_{H_u}{\bf Y}_{\nu}^{\dagger} {\bf Y}_{\nu}
+{\bf A}_{\nu}^{\dagger} {\bf A}_{\nu} \right)_i^j
\nonumber \\
&&
\left.
-\left (\frac{6}{5}g_1^2 \left| M_1 \right|^2
+6 g_2^2 \left| M_2 \right|^2 \right) \delta_i^j
\right ],
\nonumber\\
\mu \frac{d}{d \mu} ({\bf m}^2_{\tilde e})^i_j&=&
\frac{1}{16 \pi^2} \left [
2 \left({\bf m}^2_{\tilde e} {\bf Y}_{e} {\bf Y}_{e}^{\dagger}
+{\bf Y}_{e} {\bf Y}_{e}^{\dagger} {\bf m}^2_{\tilde e} \right)^i_j
\right.
\nonumber \\
&&
\left.
+4 \left( {\bf Y}_{e} {\bf m}^2_{\tilde \sL} {\bf Y}_{e}^{\dagger}
+m^2_{H_d}{\bf Y}_{e} {\bf Y}_{e}^{\dagger}
+{\bf A}_{e} {\bf A}_{e}^{\dagger}\right)^i_j
-\frac{24}{5} g_1^2 \left| M_1 \right|^2 \delta^i_j
\right],
\nonumber \\
\mu \frac{d}{d \mu} ({\bf m}^2_{\tilde \nu})^i_j&=&
\frac{1}{16 \pi^2} \left [
2 \left({\bf m}^2_{\tilde \nu} {\bf Y}_{\nu} {\bf Y}_{\nu}^{\dagger}
+{\bf Y}_{\nu}{\bf Y}_{\nu}^{\dagger} {\bf m}^2_{\tilde \nu} \right)^i_j
\right.
\nonumber \\
&&
\left.
+4 \left( {\bf Y}_{\nu} {\bf m}^2_{\tilde \sL} {\bf Y}_{\nu}^{\dagger}
+m^2_{Hu}{\bf Y}_{\nu} {\bf Y}_{\nu}^{\dagger}
+{\bf A}_{\nu} {\bf A}_{\nu}^{\dagger}\right)^i_j
\right],
\nonumber\\
\mu \frac{d}{d \mu} {\bf A}_{e}^{ij}&=&
\frac{1}{16 \pi^2} \left[ \left\{
-\frac{9}{5} g_1^2 -3 g_2^2+ 3 {\rm Tr}({\bf Y}_d^{\dagger} {\bf Y}_d)
+{\rm Tr}({\bf Y}_{e}^{\dagger} {\bf Y}_{e}) \right \} {\bf A}_{e}^{ij}
\right.
\nonumber \\
&&+2 \left\{
-\frac{9}{5} g_1^2 M_1 -3 g_2^2 M_2 + 3 {\rm Tr}({\bf Y}_d^{\dagger} {\bf A}_d)
+{\rm Tr}({\bf Y}_{e}^{\dagger} {\bf A}_{e}) \right \} {\bf Y}_{e}^{ij}
\nonumber \\
&&\left.
+4 ({\bf Y}_{e} {\bf Y}_{e}^{\dagger} {\bf A}_{e})^{ij} + 5 ({\bf A}_{e} {\bf Y}_{e}^{\dagger} {\bf Y}_{e})^{ij}
+2({\bf Y}_{e} {\bf Y}_{\nu}^{\dagger} {\bf A}_{\nu})^{ij} +
 ({\bf A}_{e} {\bf Y}_{\nu}^{\dagger} {\bf Y}_{\nu})^{ij}
\right],
\nonumber\\
\mu \frac{d}{d \mu} {\bf A}_{\nu}^{ij}&=&
\frac{1}{16 \pi^2} \left[ \left\{
-\frac{3}{5}g_1^2 -3g_2^2 +3 {\rm Tr}({\bf Y}_u^{\dagger} {\bf Y}_u)
+{\rm Tr}({\bf Y}_{\nu}^{\dagger} {\bf Y}_{\nu}) \right \} {\bf A}_{\nu}^{ij}
\right.
\nonumber \\
&&+2 \left\{
-\frac{3}{5} g_1^2 M_1 -3 g_2^2 M_2 + 3 {\rm Tr}({\bf Y}_u^{\dagger} {\bf A}_u)
+{\rm Tr}({\bf Y}_{\nu}^{\dagger} {\bf A}_{\nu}) \right \} {\bf Y}_{\nu}^{ij}
\nonumber \\
&&\left.
+4({\bf Y}_{\nu} {\bf Y}_{\nu}^{\dagger}{\bf A}_{\nu})^{ij}
+5({\bf A}_{\nu}{\bf Y}_{\nu}^{\dagger} {\bf Y}_{\nu})^{ij}
+2({\bf Y}_{\nu} {\bf Y}_{e}^{\dagger}{\bf A}_{e})^{ij}
+({\bf A} _{\nu} {\bf Y}_{e}^{\dagger} {\bf Y}_{e})^{ij}
\right],
\nonumber\\
\mu \frac{d}{d \mu} {\bf A}_{u}^{ij}&=&
\frac{1}{16 \pi^2} \left[ \left\{
-\frac{13}{15}g_1^2 -3g_2^2 -\frac{16}{3}g_3^2 +3 {\rm Tr}({\bf Y}_u^{\dagger} {\bf Y}_u)
+{\rm Tr}({\bf Y}_{\nu}^{\dagger} {\bf Y}_{\nu}) \right \} {\bf A}_{u}^{ij}
\right.
\nonumber \\
&&+2 \left\{
-\frac{13}{15} g_1^2 M_1 -3 g_2^2 M_2 -\frac{16}{3}g_3^2 M_3 
+3 {\rm Tr}({\bf Y}_u^{\dagger} {\bf A}_u)
+{\rm Tr}({\bf Y}_{\nu}^{\dagger} {\bf A}_{\nu}) \right \} {\bf Y}_{u}^{ij}
\nonumber \\
&&\left.
+4({\bf Y}_u {\bf Y}_u^{\dagger}{\bf A}_u)^{ij}
+5({\bf A}_u {\bf Y}_u^{\dagger} {\bf Y}_u)^{ij}
+2({\bf Y}_u {\bf Y}_d^{\dagger}{\bf A}_d)^{ij}
+({\bf A}_u {\bf Y}_d^{\dagger} {\bf Y}_d)^{ij}
\right],
\nonumber\\
\mu \frac{d}{d \mu} {\bf A}_{d}^{ij}&=&
\frac{1}{16 \pi^2} \left[ \left\{
-\frac{7}{15}g_1^2 -3g_2^2 -\frac{16}{3}g_3^2 +3 {\rm Tr}({\bf Y}_d^{\dagger} {\bf Y}_d)
+{\rm Tr}({\bf Y}_{e}^{\dagger} {\bf Y}_{e}) \right \} {\bf A}_{d}^{ij}
\right.
\nonumber \\
&&+2 \left\{
-\frac{7}{15} g_1^2 M_1 -3 g_2^2 M_2 -\frac{16}{3}g_3^2 M_3 
+3 {\rm Tr}({\bf Y}_d^{\dagger} {\bf A}_d)
+{\rm Tr}({\bf Y}_{e}^{\dagger} {\bf A}_{e}) \right \} {\bf Y}_{d}^{ij}
\nonumber \\
&&\left.
+4({\bf Y}_d {\bf Y}_d^{\dagger}{\bf A}_d)^{ij}
+5({\bf A}_d {\bf Y}_d^{\dagger} {\bf Y}_d)^{ij}
+2({\bf Y}_d {\bf Y}_u^{\dagger}{\bf A}_u)^{ij}
+({\bf A}_d {\bf Y}_u^{\dagger} {\bf Y}_u)^{ij}
\right],
\nonumber
\end{eqnarray}

\begin{eqnarray}
\mu \frac{d}{d \mu} (m^2_{H_u})&=&
\frac{1}{16\pi^2}
\left[
6{\rm {\rm Tr}}\left(
{\bf m}^2_{\tilde Q}{\bf Y}^{\dagger}_{u}{\bf Y}_{u}
+{\bf Y}^{\dagger}_{u}
({\bf m}_{\tilde u}
+m^2_{H_u})
{\bf Y}_{u}
+{\bf A}^{\dagger}_{u}{\bf A}_{u}
\right)
\right.
\nonumber\\
&&\left.
+2{\rm Tr}\left(
{\bf m}^2_{\tilde \sL}{\bf Y}^{\dagger}_{\nu}{\bf Y}_{\nu}
+{\bf Y}^{\dagger}_{\nu}
({\bf m}^2_{\tilde \nu}+m^2_{H_u}){\bf Y}_{\nu}
+{\bf A}^{\dagger}_{\nu}{\bf A}_{\nu}
\right)
\right.
\nonumber\\
&&\left.
-\left(
\frac{6}{5}g^{2}_{1}|M_{1}|^{2}+6g^{2}_{2}|M_{2}|^{2}
\right)
\right],
\nonumber\\
\mu \frac{d}{d \mu} (m^2_{H_d})&=&
\frac{1}{16\pi^2}
\left[
6{\rm Tr}\left(
{\bf m}^2_{\tilde Q}{\bf Y}^{\dagger}_{d}{\bf Y}_{d}
+{\bf Y}^{\dagger}_{d}
({\bf m}_{\tilde d}
+m^2_{H_d})
{\bf Y}_{d}
+{\bf A}^{\dagger}_{u}{\bf A}_{d}
\right)
\right.
\nonumber\\
&&\left.
+2{\rm Tr}\left(
m^2_{\tilde \sL}{\bf Y}^{\dagger}_{e}{\bf Y}_{e}
+{\bf Y}^{\dagger}_{e}
({\bf m}^2_{\tilde e}+m^2_{H_d}){\bf Y}_{e}
+{\bf A}_{e}^{\dagger}{\bf A}_{e}
\right)
\right.
\nonumber\\
&&\left.
-\left(
\frac{6}{5}g^{2}_{1}|M_{1}|^{2}+6g^{2}_{2}|M_{2}|^{2}
\right)
\right].
\nonumber
\end{eqnarray}

\section{Notations and Conventions in the MSSM}
\subsection{Mass matrix and mixings}
In this appendix, we give our notation of SUSY particle masses and mixings
in our calculation. 

The slepton mass ${\bf {\hat M}^2}$ term is 
\begin{eqnarray}
 &&(\tilde{e}_{\sL}^{\dag},\tilde{e}_{\sR}^{\dag})
\left(
\begin{array}{cc}
 {\bf m}_{\sL}^{2} & {\bf m}_{\sL\sR}^{2\,{\rm T}} \\
 {\bf m}_{\sL\sR}^{2} & {\bf m}_{\sR}^{2} \\
\end{array}
\right)
\left(
\begin{array}{c}
 \tilde{e}_{\sL} \\
 \tilde{e}_{\sR} \\
\end{array}
\right) \ ,
\end{eqnarray}
with
\begin{eqnarray}
({\bf m}_{\sL}^{2})_{ij}&=&
({\bf m}_{\tilde{\sL}}^{2})_{ij}+m_{e_{i}}^{2}\delta_{ij}+m_{\rm \sZ}^{2}\delta_{ij}\cos2\beta
(-\frac{1}{2}+\sin^{2}\theta_{\rm \sW})\ ,\\
({\bf m}_{\sR}^{2})_{ij}&=&
({\bf m}_{\tilde{e}_{\sR}})_{ij}+m_{e_{i}}^{2}\delta_{ij}-m_{\rm \sZ}^{2}\delta_{ij}\cos2\beta\sin^{2}\theta_{\rm \sW}\ ,\\
({\bf m}_{\sL\sR}^{2})_{ij}&=&
\frac{{\bf A}^{e}_{ij}v\cos\beta}{\sqrt{2}}-m_{e_{i}}\mu\tan\beta\ ,
\end{eqnarray}
\noindent where
$({\bf m}_{\sL}^{2})_{ij}$ and $({\bf m}_{\sR}^{2})_{ij}$ are $3\times3$ matrix.
The slepton mass matrix can be diagonalized as 
\begin{eqnarray}
 U^{f}{\bf {\hat M}^2}U^{f{\rm T}}=({\rm diagonal})\ ,
\end{eqnarray}
\noindent where
$ U^{f}$ is a real orthogonal $6\times6$  matrix.

The chargino mass term is 
\begin{eqnarray}
-{\cal L}&=&(\overline{\widetilde{W}_{\sR}^{-}},\overline{\widetilde{H}_{2\sR}^{-}})
\left(
\begin{array}{cc}
 M_{2} & \sqrt{2}m_{\rm \sW}\cos\beta\\
 \sqrt{2}m_{\rm \sW}\sin\beta & \mu 
\end{array}
\right)
\left(
\begin{array}{c}
 \widetilde{W}_{\sL}^{-} \\
 \widetilde{H}_{1\sL}^{-} \\
\end{array}
\right)+{\rm h.c.}\ .
\end{eqnarray}
The chargino mass matrix can be diagonalized as 
\begin{eqnarray}
 O_{\sR}M_{\sC}O_{\sL}^{\rm T}={\rm diag}(M_{\tilde{\chi}_{1}^{-}},M_{\tilde{\chi}_{2}^{-}})\ ,
\end{eqnarray}
\noindent where
$O_{\sL}$ and $O_{\sR}$ are real orthogonal $2\times2$ matrix. 
The mass eigenstate $\tilde{\chi}_{\sA\sL}$ and $\tilde{\chi}_{\sA\sR}$
$(A=1,2)$ are  
\begin{eqnarray}
\left(
\begin{array}{c}
 \tilde{\chi}_{1\sL}^{-} \\
 \tilde{\chi}_{2\sL}^{-}
\end{array}
\right)
=
O_{\sL}
\left(
\begin{array}{c}
 \widetilde{W}_{\sL}^{-} \\
 \widetilde{H}_{1\sL}^{-}
\end{array}
\right)\ ,\ \ \ 
\left(
\begin{array}{c}
 \tilde{\chi}_{1\sR}^{-} \\
 \tilde{\chi}_{2\sR}^{-}
\end{array}
\right)
=
O_{\sR}
\left(
\begin{array}{c}
 \widetilde{W}_{\sR}^{-} \\
 \widetilde{H}_{2\sR}^{-}
\end{array}
\right)\ ,
\end{eqnarray}
\noindent and 
\begin{eqnarray}
 \tilde{\chi}_{\sA}^{-}=\tilde{\chi}_{\sA\sL}^{-}+\tilde{\chi}_{\sA\sR}^{-}\ \ \ \ \ (A=1,2)
\end{eqnarray}
forms Dirac fermion with mass $M_{\tilde{\chi}_{A}^{-}}$.

The neutralino mass term is 
\begin{eqnarray}
-{\cal L}=\frac{1}{2}
(\widetilde{B}_{\sL},\widetilde{W}_{\sL}^{0},\widetilde{H}_{1\sL}^{0},\widetilde{H}_{2\sL}^{0})
M_{N}
\left(
\begin{array}{c}
 \widetilde{B}_{\sL} \\
 \widetilde{W}_{\sL}^{0} \\
 \widetilde{H}_{1\sL}^{0} \\
 \widetilde{H}_{2\sL}^{0} \\
\end{array}
\right)+{\rm h.c.} \ ,
\end{eqnarray}
\noindent where
\begin{eqnarray}
M_{N}=
\left(
\begin{array}{cccc}
M_{1} & 0  & -m_{\rm Z}\sin\theta_{\rm W}\cos\beta & 
m_{\rm Z}\sin\theta_{\rm W}\sin\beta \\
0     & M_{2}& m_{\rm Z}\cos\theta_{\rm W}\cos\beta &  
-m_{\rm Z}\cos\theta_{\rm W}\sin\beta \\ 
-m_{\rm Z}\sin\theta_{\rm W}\cos\beta & 
m_{\rm Z}\cos\theta_{\rm W}\cos\beta & 0 & -\mu\\
m_{\rm Z}\sin\theta_{\rm W}\sin\beta &  
-m_{\rm Z}\cos\theta_{\rm W}\sin\beta & -\mu & 0\\ 
\end{array}
\right).  \nonumber\\
\end{eqnarray}
The neutralino mass matrix can be diagonalized as
\begin{eqnarray}
 O_{\sN}M_{\sN}O_{\sN}^{\rm T}=({\rm diagonal})\ ,
\end{eqnarray}
where $O_{\sN}$ is a real $4\times4$ orthogonal matrix. 
The mass eigenstate is 
\begin{eqnarray}
 \tilde{\chi}_{\sA\sL}^{0}=(O_{\sN})_{\sA\sB}\tilde{\chi}_{\sB\sL}^{0}\ (A,B=1,\cdots,4)\ \ \ ,\ \ \ 
 \tilde{\chi}_{\sB\sL}^{0}=(\widetilde{B}_{\rm \sL},\widetilde{W}_{\rm \sL}^{0},\widetilde{H}_{\rm 1\sL}^{0},\widetilde{H}_{\rm 2\sL}^{0})
\end{eqnarray}
\noindent and 
\begin{eqnarray}
 \tilde{\chi}_{\sA}^{0}=\tilde{\chi}_{\sA\sL}^{0}+\tilde{\chi}_{\sA\sR}^{0}\ \ \ \ \ (A=1,\cdots,4) 
\end{eqnarray}
forms Mojorana spinor with mass $M_{\tilde{\chi}_{A}^{0}}$. 

The chargino vertex functions are 
\begin{eqnarray}
 C_{e\sA\sX}^{\sR(\ell)}&=&-g_{2}(O_{\sR})_{\sA1}U_{\sX,1}^{\nu}\ ,\\
 C_{e\sA\sX}^{\sL(\ell)}&=&g_{2}\frac{m_{e}}{\sqrt{2}m_{\rm \sW}\cos\beta}
(O_{\sL})_{\sA2}U_{\sX,1}^{\nu}\ ,
\end{eqnarray}
and the neutralino vertex functions are 
\begin{eqnarray}
 N_{e\sA\sX}^{R(\ell)}&=&
-\frac{g_{2}}{\sqrt{2}}
\left\{
\left[
-(O_{\sN})_{\sA2}-(O_{\sN})_{\sA1}\tan\theta_{\rm \sW}
\right]
U_{\sX,1}^{\ell}+
\frac{m_{e}}{m_{\rm \sW}\cos\beta}(O_{\sN})_{\sA3}U_{\sX,4}
\right\},\ \ \ \ \ \\
 N_{e\sA\sX}^{\sL(\ell)}&=&
-\frac{g_{2}}{\sqrt{2}}
\left\{
\frac{m_{e}}{m_{\rm \sW}\cos\beta}(O_{\sN})_{\sA3}U_{\sX,1}
-2(O_{\sN})_{\sA1}\tan\theta_{\rm \sW}U_{\sX,4}^{\ell}
\right\}\ .
\end{eqnarray}
\subsection{Decay amplitudes $A^{\sL,\sR}$}
For the amplitudes  $A^{\sL,\sR}$, 
there are the contributions of the chargino loop and the neutralino loop:
\begin{eqnarray}
 A^{\sL,\sR}=A^{(c)\sL,\sR}+A^{(n)\sL,\sR}\ .
\end{eqnarray}
The contributions from the chagino loop are 
\begin{eqnarray}
 A^{(c)\sL}&=&
-\frac{1}{32\pi^{2}}\frac{1}{m_{\tilde{\nu}_{X}}^{2}}
\left[
C_{j\sA\sX}^{\sL(\ell)}C_{i\sA\sX}^{\sL(\ell)*}
\frac{1}{6(1-x_{\sA\sX})^{4}}(2+3x_{\sA\sX}-6x^{2}_{\sA\sX}+x^{3}_{\sA\sX}+6x_{\sA\sX}\ln x_{\sA\sX})
\right.\nonumber\\
&&\left.\hspace{1.5 cm}+
C_{j\sA\sX}^{\sL(\ell)}C_{i\sA\sX}^{\sR(\ell)*}
\frac{M_{\tilde{\chi}_{\sA}^{-}}}{m_{j}}
\frac{1}{(1-x_{\sA\sX})^{3}}(-3+4x_{\sA\sX}-x^{2}_{\sA\sX}-2\ln x_{\sA\sX})
\right] ,\\
A^{(c)\sR}&=&A^{(c)\sL}|_{\sL\leftrightarrow \sR}\ ,
\end{eqnarray}
where $x_{\sA\sX}$ is defined as 
\begin{eqnarray}
x_{\sA\sX}
=
\frac{M_{\tilde{\chi}^{-}_{A}}^{2}}{m_{\tilde{\nu}_{X}}^{2}}\ .
\end{eqnarray}
Here
$m_{\tilde{\nu}_{X}}$ is the sneutrino mass and $M_{\tilde{\chi}^{-}_{A}}$ is 
the chargino mass. 
The contributions from the neutralino loop are  
\begin{eqnarray}
 A^{(n)\sL}&=&
\frac{1}{32\pi^{2}}\frac{1}{m_{\tilde{\ell}_{\sX}}^{2}}
\left[
N_{j\sA\sX}^{\sL(\ell)}N_{i\sA\sX}^{\sL(\ell)*}
\frac{1}{6(1-y_{\sA\sX})^{4}}(1-6y_{\sA\sX}+3y^{2}_{\sA\sX}+2y^{3}_{\sA\sX}-6y^{2}_{\sA\sX}\ln y_{\sA\sX})
\right.\nonumber\\
&&\left.\hspace{2.0cm}+
N_{j\sA\sX}^{\sL(\ell)}N_{i\sA\sX}^{\sR(\ell)*}
\frac{M_{\tilde{\chi}_{\sA}^{0}}}{m_{i}}
\frac{1}{(1-y_{\sA\sX})^{3}}(1+y^{2}_{\sA\sX}+2y_{\sA\sX}\ln y_{\sA\sX})
\right]\ ,\\
A^{(n)\sR}&=&A^{(n)\sL}|_{\sL\leftrightarrow \sR}\ ,
\end{eqnarray}
where $y_{\sA\sX}$ is defined as 
\begin{eqnarray}
 y_{\sA\sX}=\frac{M_{\tilde{\chi}^{0}_{A}}^{2}}{m_{\tilde{\ell}_{X}}^{2}}\ .
\end{eqnarray}
Here $m_{\tilde{\ell}_{\sX}}$ is the charged slepton mass and
$M_{\tilde{\chi}^{0}_{\sA}}$ is the neutralino mass. 
\section{Anomalous $U(1)$ flavor symmetry}

We review the anomalous $U(1)$ flavor symmetry \cite{U1} which is
utilized in the
Shafi-Tavartkiladze model \cite{Shafi} discussed in section 4. 
The anomalous $U(1)$ flavor symmetry can arise from string theory.
The cancellation of this anomaly is due to the Green-Schwarz mechanism 
\cite{Green}.
The associated Fayet-Iliopoulos term is given as \cite{Anomalous}
\begin{eqnarray}
 \xi\int d^{4}\theta V_{\sA}\ \quad
\ {\rm with }\qquad \xi=\frac{g_{\sA}^{2}M_{\rm pl}^{2}}{192\pi^{2}}{\rm Tr}Q\ .
\end{eqnarray}
The D-term is given as 
\begin{eqnarray}
 \frac{g_{\sA}^{2}}{8}D_{\sA}^{2}
=\frac{g_{\sA}^{2}}{8}
 \left(
   \sum Q_{a}|\varphi_{a}|^{2}+\xi
 \right)^{2}\ ,\label{D-term}
\end{eqnarray}
where $Q_{a}$ is the `anomalous' charge of $\varphi_{a}$.  
For $U(1)$ breaking, we introduce the singlet field $S$ 
under the SM gauge group with $U(1)$ charge
$Q_{\sS}$.
Assuming Tr$Q>0$, we can ensure the cancellation of $D_{\sA}$ in
(\ref{D-term}).
Taking $Q_{\sS}=-1$, we can ensure the non-zero VEV of S:
$\langle S\rangle$ which
is given as $\langle S\rangle=\sqrt{\xi}$.

Due to the Froggatt-Nielsen mechanism,
Yukawa interaction term in the effective theory is given as 
\begin{eqnarray}
 e^{c}_{{\sR}j}L_{i}H_{d}
\left(
\frac{S}{M_{\rm pl}}
\right)^{m_{ij}} \ ,
\label{YukawaInt1}
\end{eqnarray}
where $e^{c}_{\sR{j}}$ and $L_{i}$ are the right-handed charged lepton and
left-handed lepton doublet, respectively, 
$H_{d}$ is Higgs doublet, and 
$S$ is singlet field.
The effective Yukawa couplings are given in terms of 
\begin{eqnarray}
 \lambda\equiv\frac{\langle S\rangle}{M_{\rm pl}} \ .
\end{eqnarray}

In order to make  the interaction term eq.(\ref{YukawaInt1}) neutral, 
   Shafi and Tavartkiladze 
assigned $U(1)$ flavor charges as follows:
\begin{eqnarray}
 Q_{\rm \sL_{1}}=k+n\ ,\  Q_{\rm \sL_{2}}= Q_{\rm \sL_{3}}=k,\ 
 Q_{\rm \sN_{1}}= -Q_{\rm \sN_{2}}=k+k^{'},\nonumber\\ 
 Q_{H_{u}}=Q_{H_{d}}=0,\ \ Q_{\sS}=-1\ ,\hspace{2.5 cm}
\end{eqnarray}
where $k,n,k^{'}>0,\ n\geq k^{'}$.
Then,  they obtained
 
\begin{eqnarray}
  {\bf Y}_{\nu}&=&
\left(
\begin{array}{ccc}
 \lambda^{2k+n+k^{'}}& \lambda^{2k+k^{'}}& \lambda^{2k+k^{'}}\\
 \lambda^{n-k} & 0 & 0
\end{array}
\right) \ ,
\end{eqnarray}
\noindent and 
\begin{eqnarray}
 {\bf M}_{\bf\sR}&=&
M_{\sR}
\left(
\begin{array}{cc}
 \lambda^{2k+2k^{'}}& 1\\
 1& 0 
\end{array}
\right) \ .
\end{eqnarray}
In conculsion, the neutrino mass matrix is given as  
\begin{eqnarray}
 {\bf m}_{\nu}=
{\bf Y}_{\nu}^{\rm T}{\bf M}_{\bf\sR}^{-1}{\bf Y}_{\nu}v_{u}^{2}=
\frac{\lambda^{2k+n}v_{u}^{2}}{M_{\sR}}
\left(
\begin{array}{ccc}
\lambda^{n} & 1& 1\\
1& 0&0\\
1&0&0
\end{array}
\right) \ .
\end{eqnarray}

\end{document}